\documentclass[graybox]{svmult}

\usepackage{mathptmx}       
\usepackage{graphicx}        
\usepackage[bottom]{footmisc}

\begin{document}

\title*{The Kuiper Belt and Other Debris Disks}

\author{David Jewitt, Amaya Moro-Mart\'{\i}n and Pedro Lacerda}
\institute{David Jewitt and Pedro Lacerda \at Institute for Astronomy,
University of Hawaii, \email{jewitt@ifa.hawaii.edu} \and Amaya Moro-Mart\'{\i}n
\at Dept. Astronomy, Princeton University \email{amaya@astro.princeton.edu}}
\maketitle

\abstract{We discuss the current knowledge of the Solar system, focusing on
bodies in the outer regions, on the information they provide concerning Solar
system formation, and on the possible relationships that may exist between our
system and the debris disks of other stars.  Beyond the domains of the
Terrestrial and giant planets, the comets in the Kuiper belt and the Oort cloud
preserve some of our most pristine materials.  The Kuiper belt, in particular,
is a collisional dust source and a scientific bridge to the dusty ``debris
disks'' observed around many nearby main-sequence stars.  Study of the Solar
system provides a level of detail that we cannot discern in the distant disks
while observations of the disks may help to set the Solar system in proper
context. }

\section{Introduction}

Planetary astronomy is unusual among the astronomical sciences in that the
objects of its attention are inexorably transformed by intensive study into the
targets of other sciences.   For example, the Moon and Mars were studied
telescopically by astronomers for nearly four centuries but, in the last few
decades, these worlds have been transformed into the playgrounds of geologists,
geophysicists, meteorologists, biologists and others.  Telescopic studies
continue to be of value, but we now learn most about the Moon and Mars from
in-situ investigations.  This transformation from science at-a-distance to
science up-close is a forward step and a tremendous luxury not afforded to the
rest of astronomy.  Those who study other stars or the galaxies beyond our own
will always be forced by distance to do so telescopically.  

However, the impression that the Solar system is now \textit{only} geology or
meteorology or some other science beyond the realm of astronomy is completely
incorrect when applied to the outer regions.  The Outer Solar system (OSS)
remains firmly entrenched within the domain of astronomy, its contents
accessible only to telescopes. Indeed, major components of the OSS, notably the
Kuiper belt, were discovered (telescopically) less than two decades ago and
will continue to be best studied via. astronomical techniques for the
foreseeable future. It is reasonable to expect that the next generation of
telescopes in space and on the ground will play a major role in improving our
understanding of the Solar system, its origin and its similarity to related
systems around other stars.  

In this chapter, we present an up-to-date overview of the layout of the Solar
system and direct attention to the outer regions where our understanding is the
least secure but the potential for scientific advance is the greatest.   The
architecture of the Kuiper belt is discussed as an example of a source-body
system that probably lies behind many of the debris disks of other stars.
Next, the debris disks are discussed based on the latest observations from the
ground and from Spitzer, and on new models of dust transport.  Throughout, we
use text within grey boxes to highlight areas where the Next Generation
telescopes are expected to have major impact.

\section{The Architecture of the Solar System}
\label{Intro}

It is useful to divide the Solar system into three distinct domains, those of
the Terrestrial planets, the giant planets, and the comets.  Objects within
these domains are distinguished by their compositions, by their modes of
formation and by the depth and quality of knowledge we possess on each.  

\subsection{Terrestrial Planets}

The Terrestrial planets (Mercury, Venus, Earth and Moon, Mars and most
main-belt asteroids) are found inside  3 AU.  They have refractory compositions
dominated by iron ($\sim$ 35\% by mass), oxygen ($\sim$ 30\%), silicon and
magnesium ($\sim$15\% each) and were formed by binary accretion in the
protoplanetary disk.   About 95\% of the Terrestrial planet mass is contained
within Venus and Earth ($\sim$1 M$_{\oplus}$ = 6$\times$10$^{24}$ kg, each).
The rest is found in the small planets Mercury and Mars, with trace amounts
($\sim$3$\times$10$^{-4}$ M$_{\oplus}$) in the main-belt asteroids located
between Mars and Jupiter.  The largest asteroid is Ceres ($\sim$ 900 km
diameter).  The Terrestrial planets grew by binary accretion between solid
bodies in the protoplanetary disk of the Sun.  While not all details of this
process are understood, it is clear that sticking and coagulation of dust
grains, perhaps aided at first by hydrodynamic forces exerted from the gaseous
component of the disk, produced larger and larger bodies up to the ones we see
now in the Solar system.  The gaseous component, judged mainly by observations
of other stars, dissipated on timescales from a few to $\sim$10 Myr.
Measurements of inclusions within primitive meteorites show that macroscopic
bodies existed within a few Myr of the origin.  The overall timescale for
growth was determined, ultimately, by the sweeping up of residual mass from the
disk, a process thought to have taken perhaps 40 Myr in the case of the Earth.
No substantial body is found in the asteroid belt although it is likely that
sufficient mass existed there to form an object of planetary class.  This is
thought to be because growth  in this region was interrupted by strong
perturbations, caused by the emergence of nearby, massive Jupiter (at $\sim$5
AU).  
 
\subsection{Giant Planets}

Orbits of the giant planets (Jupiter, Saturn, Uranus and Neptune) span the
range 5 AU to 30 AU.  The giants are in fact of two compositionally distinct
kinds.  

\subsubsection{Gas Giants}

Jupiter (310 M$_{\oplus}$) and Saturn (95 M$_{\oplus}$) are so-called because,
mass-wise, they are dominated by hydrogen and helium.  Throughout the bulk of
each planet these gases are compressed, however, into a degenerate (metallic)
liquid that supports convection and sustains a magnetic field through dynamo
action. The compositional similarity to the Sun suggests to some investigators
that the gas giants might form by simple hydrodynamic collapse of the
protoplanetary gas nebula (Boss 2001).  In hydrodynamic collapse the essential
timescale is given by the free-fall time, and this could be astonishingy short
(e.g. 1000 yrs).  Details of this instability, especially related to the
necessarily rapid cooling of the collapsing planet,  remain under discussion
(Boley et al. 2007, Boss 2007).  In fact, measurements of the moment of
inertia, coupled with determinations of the equation of state of
hydrogen-helium mixtures at relevant pressures and temperatures, show that
Saturn (certainly) and Jupiter (probably) have distinct cores containing 5
M$_{\oplus}$ to 15 M$_{\oplus}$ of heavy elements (the case of Jupiter is less
compelling than Saturn because of its greater mass and central hydrostatic
pressure, leading to larger uncertainties concerning the self-compressibility
of the gas).   The presence of a dense core is the basis for the model of
formation through `` nucleated instability'', in which the core grows by binary
accretion in the manner of the Terrestrial planets, until the gravitational
escape speed from the core becomes comparable to the thermal speed of molecules
in the gas nebula.  Then, the core traps gas directly from the nebula, leading
to the large masses and gas-rich compositions observed in Jupiter and Saturn.  

Historically, the main sticking point for nucleated instability models has been
that the cores must grow to critical size \textit{before} the surrounding gas
nebula dissipates (i.e. 5 M$_{\oplus}$ to 15 M$_{\oplus}$ cores must grow in
much less than 10 Myr).  The increase in the disk surface density due to the
freezing of water as ice outside the snow-line is one factor helping to
decrease core growth times.  Another may be the radial jumping motion of the
growing cores, driven by angular momentum and energy exchange with
planetesimals (Rice and Armitage 2003).  In recent times, a consensus appears
to have emerged that Jupiter's core, at least, could have grown by binary
accretion from a disk with $\Sigma \sim$ 50 to 100 kg m$^{-2}$ on timescales
$\sim$1 Myr (Rice and Armitage 2003).  The collapse of nebular gas onto the
core after this would have been nearly instantaneous.

Recent data show that the heavier elements (at least in Jupiter, the better
studied of the gas giants) are enriched relative to hydrogen in the Sun by
factors of $\sim$2 to 4 (Owen et al. 1999), so that wholesale nebular collapse
cannot be the whole story (and may not even be part of it).  The enrichment
applies not only to species that are condensible at the $\sim$100 K
temperatures appropriate to Jupiter's orbit, but to the noble gases Ar, Kr and
Xe, which can only be trapped in ice at much lower temperatures, $<$30 K (e.g.
Bar-Nun et al. 1988) .  Therefore, Owen et al.  suggest that the gas giants
incorporate substantial mass from a hitherto unsuspected population of
ultracold bodies, presumably originating in the outer Solar system.  In a
variant of this model, cold grains from the outer Solar system trap volatile
gases but drift inwards under the action of gas drag, eventually reaching the
inner nebula when they evaporate in the heat of the Sun and enrich the gas
(Guillot and Huesco 2006).  

\subsubsection{Ice Giants}

Uranus (15 M$_{\oplus}$) and Neptune (17 M$_{\oplus}$), in addition to being an
order of magnitude less massive than the gas giants are compositionally
distinct. These planets contain a few M$_{\oplus}$ of H and He, and a much
larger fraction of the ``ices'' H$_2$O, CH$_4$ and NH$_3$.  For this reason
they are known as ``ice giants'', but the name is misleading because they are
certainly not solid bodies but are merely composed of molecules which, if they
were much colder, would be simple ices.  In terms of their mode of formation,
the difference between the ice giants and the gas giants may be largely one of
timescale.  It is widely thought that the ice giants correspond to the heavy
cores of Jupiter and Saturn, but with only vestigial hydrogen/helium envelopes
accreted from the rapidly dissipating gaseous component of the protoplanetary
disk.  

While qualitatively appealing, forming Uranus and (especially) Neptune on the
10 Myr timescale associated with the loss of the gas disk has been a major
challenge to those who model planetary growth.  The problem is evident from a
simple consideration of the collision rate between particles in a disk or
surface density $\Sigma(R)$ kg m$^{-2}$, where $R$ is the heliocentric
distance.  The probability of a collision in each orbit varies in proportion to
$\Sigma(R)$, while the orbital period varies as $R^{3/2}$.  Together, this
gives a collision timescale varying as $R^{3/2}$/$\Sigma(R)$ which, with
$\Sigma(R) \propto R^{-3/2}$ gives $t_c \propto R^{3}$.  A giant planet core
that takes 1 Myr to form at 5 AU would take 6$^3\sim$200 Myr to form at 30 AU,
and this considerably exceeds the gas disk lifetime.  Suggested solutions to
the long growth times for the outer planets include augmentation of the disk
density, $\Sigma(R)$, perhaps through the action of aerodynamic drag, and
formation of the ice giants at smaller distances (and therefore higher
$\Sigma$) than those at which they reside.   The latter possibility ties into
the general notion that the orbits of the outermost three planets have expanded
in response to the action or torques between the planets and the disk.  Still
another idea is that Uranus and Neptune are gas giants whose hydrogen and
helium envelopes were ablated by ionizing radiation from the Sun or a nearby,
hot star (Boss et al. 2002).

\subsection{Comets}

Comets are icy bodies which sublimate in the heat of the Sun, producing
observationally diagnostic unbound atmospheres or ``comae''.  For most known
comets, the sublimation is sufficiently strong that mass loss cannot be
sustained for much longer than $\sim$10$^4$ yr, a tiny fraction of the age of
the Solar system.  For this reason, the comets must be continually resupplied
to the planetary region from one or more low temperature reservoirs, if their
numbers are to remain in steady state.  In the last half century,  at least
three distinct source regions have been identified.

\subsubsection{Oort Cloud Comets}

\begin{figure}[]
\begin{center}
\includegraphics[width=0.8\textwidth]{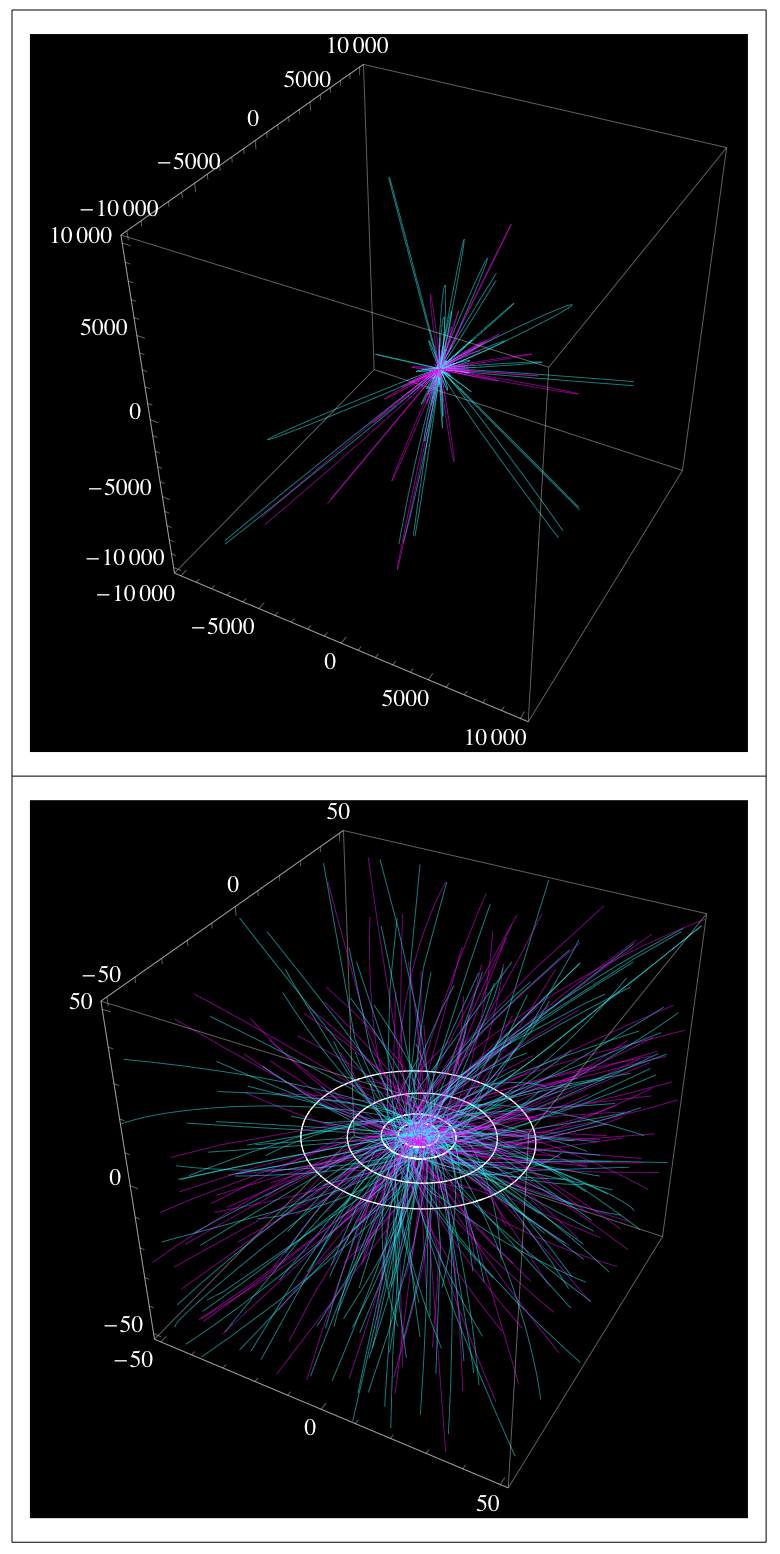}
\caption{Orbits of the nearly 200 long period comets (orbital period, $P>200$
yrs) listed in the JPL small-body database. The highly eccentric orbits of many
LPCs appear as nearly radial lines on the top (wide angle) view, showing a cube
20,000 AU on a side.  The bottom panel shows a narrow angle view of a 100 AU
cube. Prograde orbits are shown in cyan while retrograde orbits are in magenta.
Numbers along the axes are distances in AU from the Sun at (0,0,0). The orbits
of the giant planets are shown in white.}
\label{lpcs}       
\end{center}
\end{figure}

The orbits of long period comets are highly elliptical, isotropically
distributed and typically large (Figure \ref{lpcs}), suggesting a
gravitationally bound, spheroidal source region of order 100,000 AU in extent
(Oort 1950).   Comets in the cloud are scattered randomly into the planetary
region by the action of passing stars and also perturbed by the asymmetric
gravitational potential of the galactic disk (e.g. Higuchi et al. 2007).
Unfortunately, the cloud is so large that its residents cannot be directly
counted.  The population must instead be inferred from the rate of arrival of
comets from the cloud and estimates of the external torques.  A recent work
gives the number larger than $\sim$1 km in radius as 5$\times$10$^{11}$, with a
combined mass in the range 2M$_{\oplus}$ to 40M$_{\oplus}$ (Francis 2005).

\begin{figure}[]
\begin{center}
\includegraphics[width=\textwidth]{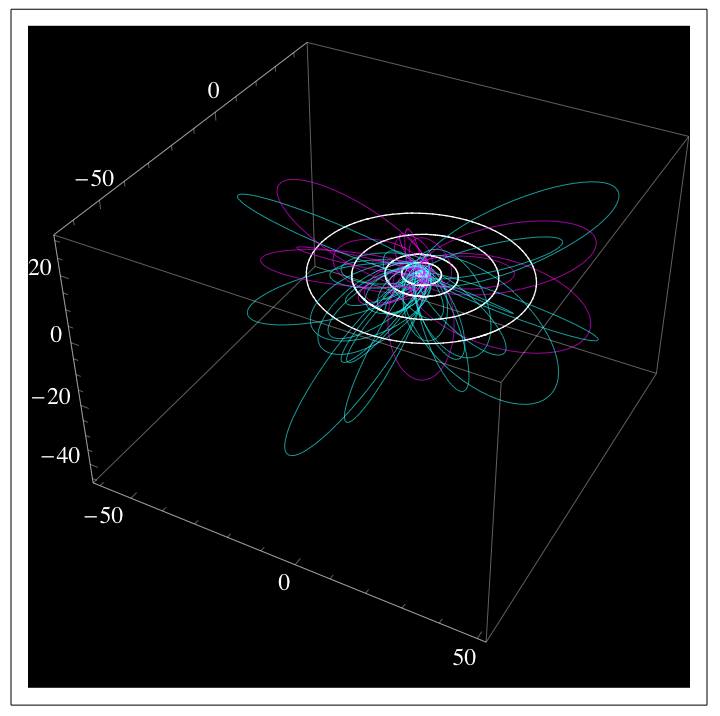}
\caption{Halley family comets.  Their non-isotropic inclination distribution,
with more prograde (cyan) orbits than retrograde (magenta), points to an origin
in the inner Oort cloud, where stellar and galactic perturbations have been too
small to randomize the orbits.  Following the classical definition, we plot the
44 comets in the JPL small-body database having Tisserand parameters with
respect to Jupiter $\le$2 and orbital periods $20<P\mathrm{(yrs)}<200$.  The
orbits of the giant planets are shown in white.
}
\label{hfcs}       
\end{center}
\end{figure}

Oort cloud comets are thought to have originated in the Sun's protoplanetary
disk in the vicinity of the giant planets and were scattered out by
interactions with the growing, migrating planetary embryos.  Although most were
lost, a fraction of the ejected comets, perhaps from 1\% to 10\%, was
subsequently deflected by external perturbations that lifted the perihelia out
of the planetary region, effectively decoupling these comets from the rest of
the system (Hahn and Malhotra 1999).  Over $\sim$1 Gyr, the orbits of trapped
comets were randomized, converting their distribution from a flattened one
reflecting their disk source to a spherical one, compatible with the random
directions of arrival of long period comets.  In this model, which is
qualitatively unchanged from that proposed by Oort (1950), the scale and
population of the Oort cloud are set by the external perturbations from nearby
stars and the galactic disk.  If the Sun formed in a cluster, then the average
perturbations from cluster members would have been bigger than now and a
substantial population of more tightly bound, so-called ``inner Oort cloud''
comets could have been trapped.  It is possible that the Halley family comets,
whose orbits are predominantly but not exclusively prograde (Figure \ref{hfcs})
are delivered from the inner Oort cloud (Levison et al. 2001).

Some 90\% to 99\% of the comets formed in our system eluded capture and now
roam the interstellar medium.  If this fraction applies to all stars, then
$\sim$10$^{23}$ to $\sim$10$^{24}$ ``interstellar comets'' exist in our galaxy,
containing several $\times$10$^5$ M$_{\odot}$ of metals.  Interstellar comets
ejected from other stars might be detected with all-sky surveys in the Pan
STARRS or Large Synoptic Survey Telescope (LSST) class (Jewitt 2003).  Such
objects would be recognized by their strongly hyperbolic orbits relative to the
Sun, quite different from any object yet observed.

\subsubsection{Kuiper Belt Comets}

\begin{figure}[]
\begin{center}
\includegraphics[width=\textwidth]{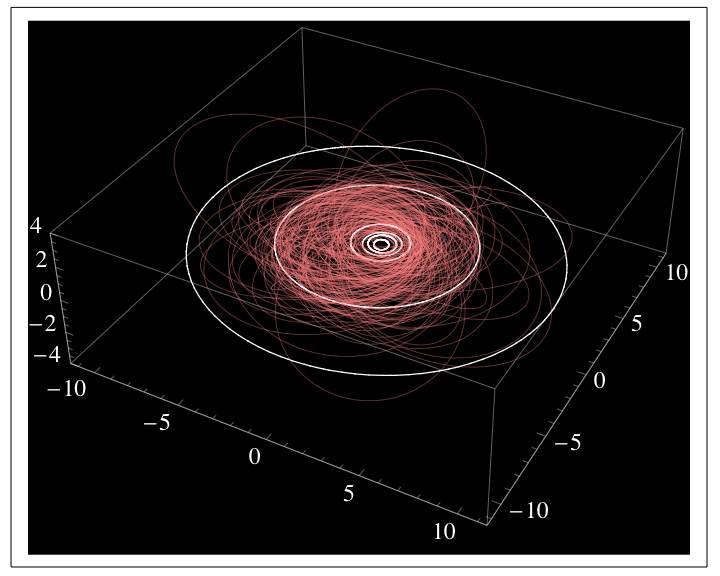}
\caption{Perspective view of the Jupiter family comets (salmon) together with
the orbits of the planets out to Saturn. The Sun is at $(x,y,z)=(0,0,0)$ and
the distances are in AU. Plotted are 166 JFCs, selected from the JPL small-body
database based on their Tisserand parameter with respect to Jupiter ($2<T_J<3$;
see, e.g., Levison and Duncan 1994) and with orbital periods $P<20$ yr.}
\label{jfcs}       
\end{center}
\end{figure}

The orbits of the Jupiter family comets (JFCs) have modest inclinations, with
no retrograde examples, and most have eccentricities much less than unity
(Figure \ref{jfcs}).  They are dynamically distinct from the long period Oort
comets, and they interact strongly with Jupiter.  For many years it was thought
that the JFCs were captured from the long period population by Jupiter but
increasingly numerical detailed work in the 1980's showed that this was not
possible (Fern\'andez 1980, Duncan et al. 1988).  The source appears to be the
Kuiper belt, although this is not an iron-clad conclusion and the particular
region or regions in the Kuiper belt from which the JFCs originate has yet to
be identified.  

About 200 JFCs are numbered (meaning that their orbits are very well
determined) and a further 200 are known.  Their survival is limited by a
combination of volatile depletion, ejection from the Solar system, or impact
into a planet or the Sun.  Dynamical interactions alone give a median lifetime
near 0.5 Myr (Levison and Duncan 1994), which matches the volatile depletion
lifetime for bodies smaller than $r\sim40$ km. The implication is that JFCs
smaller than this size will become dormant before they are dynamically removed,
ending up as bodies that are asteroidal in appearance but cometary in orbit.
Some of these dead comets are suspected to exist among the near-Earth
``asteroid'' population; indirect evidence for a fraction $\sim10\%$ comes from
albedo measurements (Fern{\'a}ndez et al. 2001) and dynamical models (Bottke et
al.  2002).

\subsubsection{Main Belt Comets}

\begin{figure}[]
\begin{center}
\includegraphics[width=\textwidth]{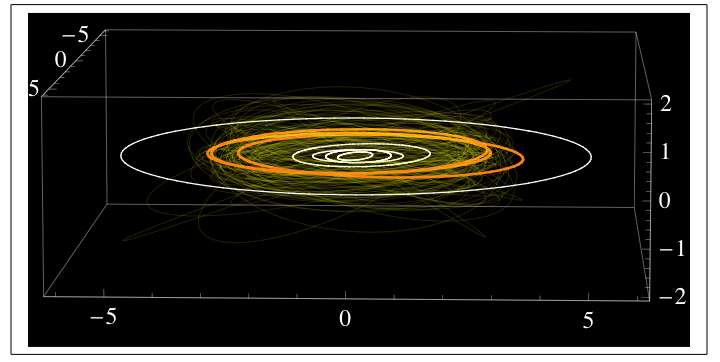}
\caption{Perspective view of the main-belt comets (orange) together with the
orbits of 100 asteroids (thin, yellow lines) and planets out to Jupiter. }

\label{mbcs}       
\end{center}
\end{figure}

\begin{figure}[]
\begin{center}
\includegraphics[width=\textwidth]{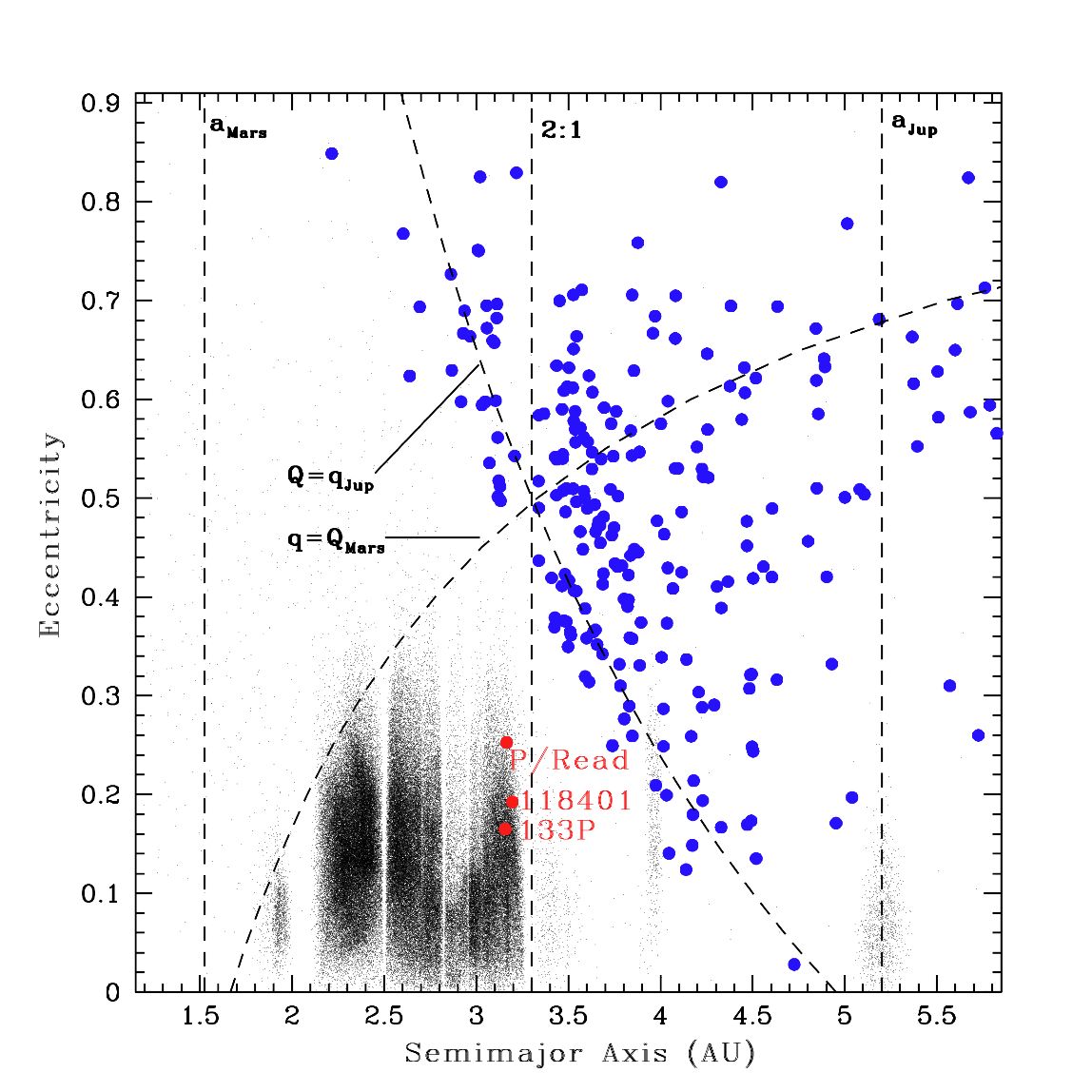}
\caption{Orbital semimajor axis vs. eccentricity for objects classified as
asteroids (black) and comets (blue), together with the three known main-belt
comets 133P/Elst-Pizarro, P/2005 U1 (Read), and 118401 (1999 RE$_{70}$)
(plotted in red). Vertical dashed lines mark the semimajor axes of Mars and
Jupiter and the 2:1 mean-motion resonance with Jupiter (commonly considered the
outer boundary of the classical main belt), as labeled.  Curved dashed lines
show the loci of orbits with perihelia equal to Mars's aphelion ($q=Q_{Mars}$)
and orbits with aphelia equal to Jupiter's perihelion ($Q=q_{Jup}$).  Objects
plotted above the $q>Q_{Mars}$ line are Mars-crossers.  Objects plotted to the
right of the $Q<q_{Jup}$ line are Jupiter-crossers.  From Hsieh and Jewitt
(2006). }

\label{mbcae}       
\end{center}
\end{figure}

At the time of writing (February 2008) three objects are known to have the
dynamical characteristics of asteroids but the physical appearances of comets.
They show comae and particle tails indicative of on-going mass loss (Hsieh and
Jewitt, 2006; Figures \ref{mbcs} and \ref{mbcae}).  These are the Main-Belt
comets (MBCs), most directly interpreted as ice-rich asteroids.   

In the modern Solar system, the MBCs are dynamically isolated from the Oort
cloud and Kuiper belt reservoirs (i.e. they cannot be captured from these other
regions given the present-day layout of the Solar system, c.f. Levison et al.
2006).  The MBCs should thus be regarded as a third and independent comet
reservoir.  Two possibilities for their origin seem plausible.  The MBCs could
have accreted ice if they grew in place but outside the snow-line.
Alternatively, the MBCs might have been captured from elsewhere if the layout
of the Solar system were very different in the past, providing a dynamical
paths from the Kuiper belt that do not now exist.   Insufficient evidence
exists at present to decide between these possibilities.

These bodies escaped detection until now because their mass loss rates ($\ll$ 1
kg s$^{-1}$) are two to three orders of magnitude smaller than from typical
comets.  The mass loss is believed to be driven by the sublimation of water ice
exposed in small surface regions, with effective sublimating areas of
$\sim$1000 m$^{-2}$ and less.  Although sublimation at 3 AU is weak, the MBCs
are small and mass loss at the observed rates cannot be sustained for the age
of the Solar system.  Instead, a trigger for the activity (perhaps impact
excavation of otherwise buried ices) is needed.  This raises the possibility
that the detected MBCs are a tiny fraction of the total number that will be
found by dedicated surveys like \textit{Pan STARRS} and \textit{LSST}. Even more interesting is
the possibility that many or even most outer belt asteroids are in fact
ice-rich bodies which display only transient activity.   Evidence from the
petrology and mineralogy of meteorites (e.g. the CI and CM chondrites that are
thought to originate in the outer belt) shows the past presence of liquid water
in the meteorite parent bodies.  The MBCs show that some of this water survived
to the present day.

\begin{svgraybox}\\

\textbf{Next generation} facilities will offer opportunities to study the physical
properties of the satellite systems of the giant planets.  The regular
satellites occupy prograde orbits of small inclination and eccentricity,
resulting from their formation in the accretion disks through which the planets
grew.  As such, the individual regular satellite systems offer information
about the sub-nebulae that must have existed around the growing planets.  These
sub-nebulae had their own density and temperature structures very different
from the local disk of the Sun.  Studies of reflected and thermal radiation
with \textit{JWST} and \textit{ALMA} will be able to determine compositional differences in even
the fainter regular satellites of the ice giants, and may provide constraints
on Io-like and Enceladus-like endogenous activity. 

The study of comets will be greatly advanced by \textit{JWST} spectroscopy of the nuclei
of comets (in order to minimize the effects of scattering from the dust coma,
these small objects must be observed when far from the Sun and therefore
faint).  Simultaneous measurements of reflected and thermal radiation will
provide albedo measurements which, with high signal-to-noise ratio spectra,
will help determine the nature and evolution of the refractory surface mantles
of these bodies.

\end{svgraybox}

\section{Kuiper Belt}

More than 1200 Kuiper belt objects (KBOs) have been discovered since the first
example, 1992 QB1, was identified in 1992.  The known objects have typical
diameters of 100 km and larger.  The total number of such objects, scaled from
published surveys, is of order 70,000, showing that there is considerable
remaining discovery space to be filled by future surveys.  The number of
objects larger than 1 km may exceed 10$^8$.

\begin{figure}[]
\begin{center}
\includegraphics[width=\textwidth]{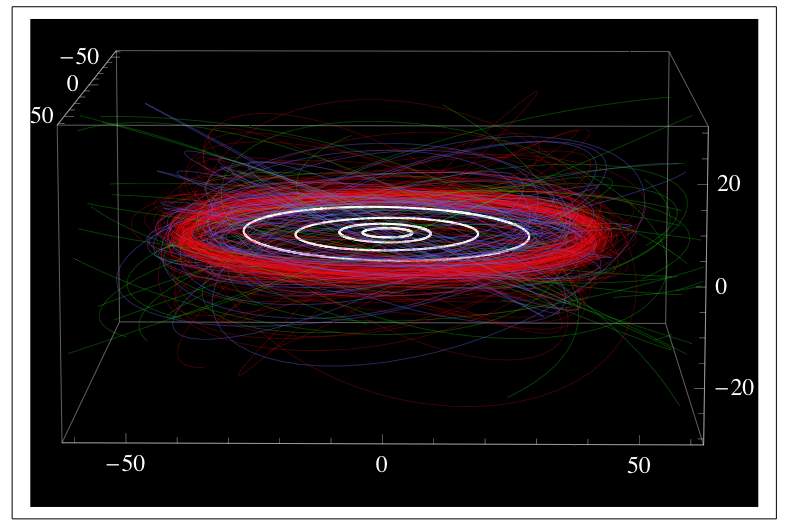}
\caption{Perspective view of the Kuiper belt.  The Sun is at $(x,y,z)=(0,0,0)$
and the axes are marked in AU.  White ellipses show the orbits of the four
giant planets.  Red orbits mark the classical KBOs.  Their ring-like
distribution is obvious.  Blue orbits show resonant KBOs, while green shows the
Scattered objects. For clarity, only one fifth of the total number of objects
in each dynamical type is plotted.  }
\label{kurious3}       
\end{center}
\end{figure}

\begin{figure}[]
\begin{center}
\includegraphics[width=\textwidth]{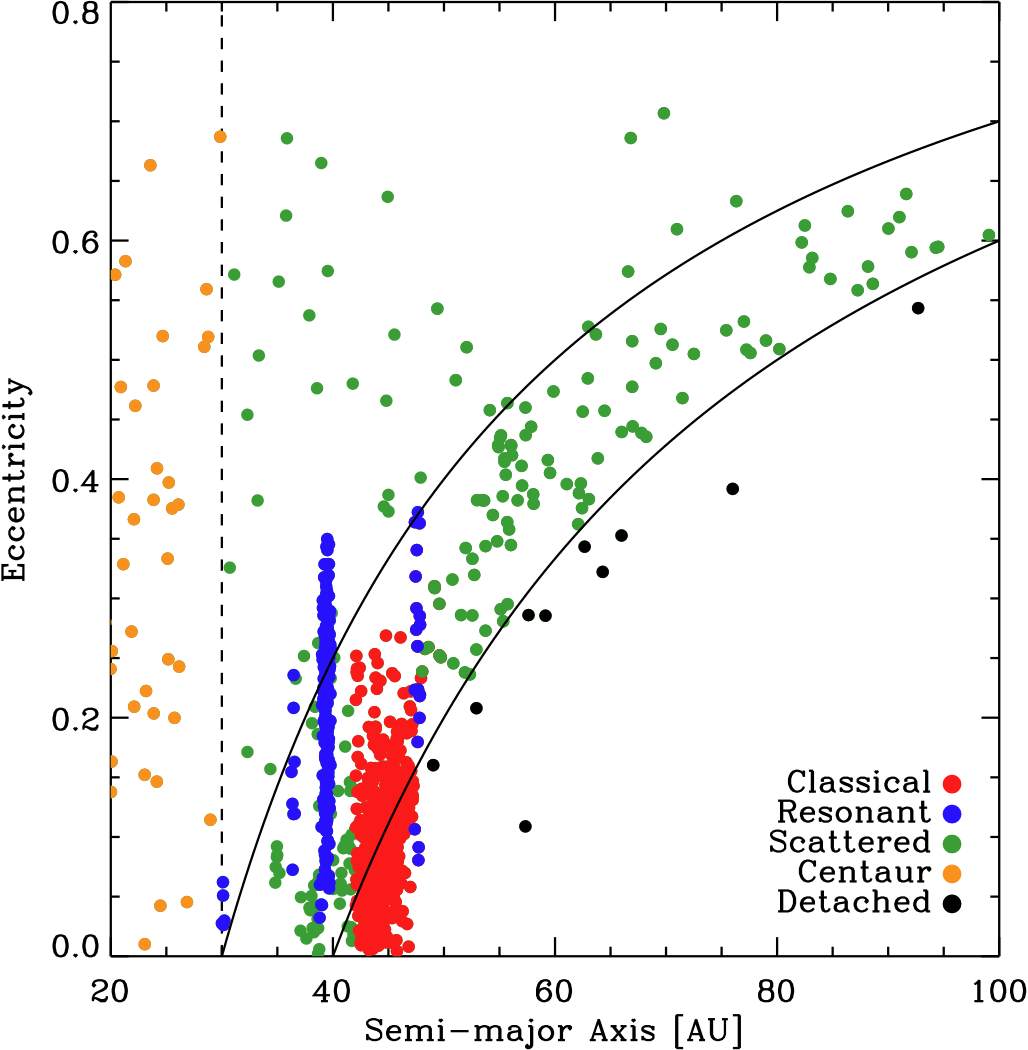}
\caption{Orbital semimajor axis vs. eccentricity for KBOs known as of 2008
January 30.  The objects are color-coded according to their dynamical type, as
labeled in the Figure and discussed in the text.  A vertical dashed line at $a$
= 30 AU marks the orbit of Neptune, the nominal inner-boundary of the Kuiper
belt.  Upper and lower solid arcs show the loci of orbits having perihelion
distances $q$ = $a(1 - e)$ equal to 30 AU and 40 AU, respectively.  }

\label{plotae}       
\end{center}
\end{figure}

The KBO orbits occupy a thick disk or sheet outside Neptune's orbit (Figure
\ref{kurious3}). A major observational result is the finding that the KBOs can
be divided on the basis of their orbital elements into several, distinct groups
(Jewitt et al. 1998).  This is best seen in $a$ (semimajor axis) vs. $e$
(orbital eccentricity) space, shown here in Figure \ref{plotae}.  The major
Kuiper belt dynamical groups in the figure are

\begin{itemize}

\item Classical Kuiper belt objects (red circles).  These orbit between about
$a$ = 42 AU and the 2:1 mean-motion resonance with Neptune at $a$ = 47 AU.
Typical orbital eccentricities of the Classicals are $e \sim$ 0 to 0.2 while
the inclination distribution appears to be bimodal with components near $i$ =
2$^{\circ}$ (so-called ``Cold-Classicals) and $i$ = 20$^{\circ}$ (the ``Hot
Classicals'', Brown and Trujillo 2001, Elliot et al. 2005).  Numerical
integrations show that the orbits of the Classical objects are stable on
timescales comparable to the age of the Solar system largely because their
perihelia are always so far from Neptune's orbit that no strong scattering
occurs.  \\

\item Resonant Kuiper belt objects (blue circles).  A number of mean-motion
resonances (MMRs) with Neptune are populated by KBOs, especially the 3:2 MMR at
39.3 AU and the 2:1 MMR at 47.6 AU (see Figure \ref{plotae}).  The 3:2 MMR
objects are known as Plutinos, to recognize 134340 Pluto as the first known
member of this population.  The 2:1 MMR objects are sometimes called
``twotinos'' while those in 1:1 MMR are Neptune's Trojans (Sheppard and
Trujillo 2006).  The resonant objects are dynamically stable by virtue of phase
protections conferred by the resonances.  For example, KBOs in 3:2 MMR can,
like Pluto itself, have perihelia inside Neptune's orbit, but their orbits
librate under perturbations from that planet in such a way that the distance of
closest approach to Neptune is always large.  In fact, all objects above the
upper solid arc in Figure \ref{plotae} have perihelia interior to Neptune's
orbit.  

The process by which KBOs became trapped in MMRs is thought to be planetary
migration (Fern{\'a}ndez and Ip 1984).  Migration occurs as a result of angular
momentum transfer during gravitational interactions between the planets and
material in the disk.  At late stages in the evolution of the disk (i.e. later
than $\sim$10 Myr, after the gaseous component of the disk has dissipated) the
interactions are between the planets and individual KBOs or other
planetesimals.  In a one-planet system, the sling-shot ejection of KBOs to the
interstellar medium would result in net shrinkage of the planetary orbit.  In
the real Solar system, however, KBOs can be scattered inwards from planet to
planet, carrying energy and angular momentum with them as they go.  Massive
Jupiter then acts as the source of angular momentum and energy, ejecting KBOs
from the system.  In the process, its orbit shrinks, while those of the other
giant planets expand.  The timescale is the same as the timescale for planetary
growth, and the distance through which a planet migrates depends on the mass
ejected from the system.  Outward migration of Neptune carries that planet's
MMRs outwards, leading to the sweep-up of KBOs (Malhotra 1995).\\

\begin{figure}[]
\begin{center}
\includegraphics[width=\textwidth]{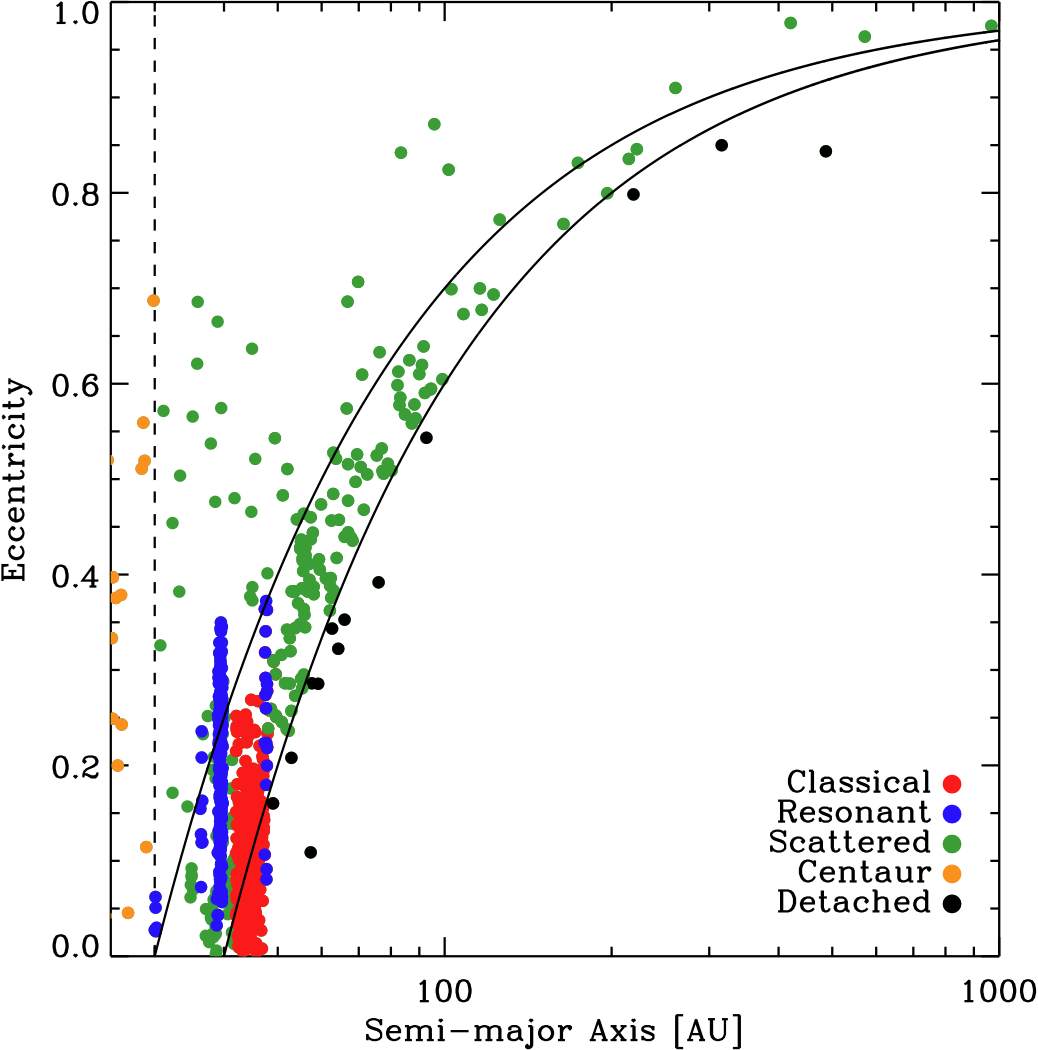}
\caption{Same as Figure \ref{plotae} but with a logarithmic x-axis extended to
$a$ = 1000 AU to better show the extent of the scattered KBOs.}

\label{plotaexl}       
\end{center}
\end{figure}

\item Scattered Kuiper belt objects (green circles), also known as Scattered
disk objects.  These objects have typically eccentric and inclined orbits with
perihelia in the 30 $\le q \le$ 40 AU range (Figures \ref{plotae} and
\ref{plotaexl}).  The prototype is 1996 TL66, whose orbit stretches from
$\sim$35 AU to $\sim$130 AU (Luu et al 1997).  The brightness of these objects
varies strongly around the orbit, such that a majority are visible only when
near perihelion.  For example, the survey in which 1996 TL66 was discovered
lacked the sensitivity to detect the object over 88\% of the orbital period.
Accordingly, the estimated population is large, probably rivaling the rest of
the Kuiper belt (Trujillo et al. 2001).  

Scattered KBOs owe their extreme orbital properties to continued, weak
perihelic interactions with Neptune which excite the eccentricity to larger and
larger values  (Duncan and Levison 1997).  The current aphelion record-holder
is 87269 (2000 OO67) with $Q$ = 1123 AU.  As the aphelion grows, so does the
dynamical influence of external perturbations from passing stars and the
asymmetric gravitational potential of the galactic disk.  

\item Detached Kuiper belt objects (black circles).  The orbits of these bodies
resemble those of the Scattered KBOs except that the perihelia are too far from
Neptune, $q >$ 40 AU, for planetary perturbations to have excited the
eccentricities.   The prototype is 2000 CR105, with $q$ = 44 AU (Gladman et al.
2002) but a more extreme example is the famous Sedna with $q$ = 74 AU (Brown et
al. 2004).  

The mechanism by which the perihelia of the detached objects were lifted away
from the influence of Neptune is unknown.  The most interesting conjectures
include the tidal action of external perturbers, whether they be unseen planets
in our own system or unbound passing stars (Morbidelli and Levison 2004).\\

\end{itemize}

\begin{figure}[]
\begin{center}
\includegraphics[width=\textwidth]{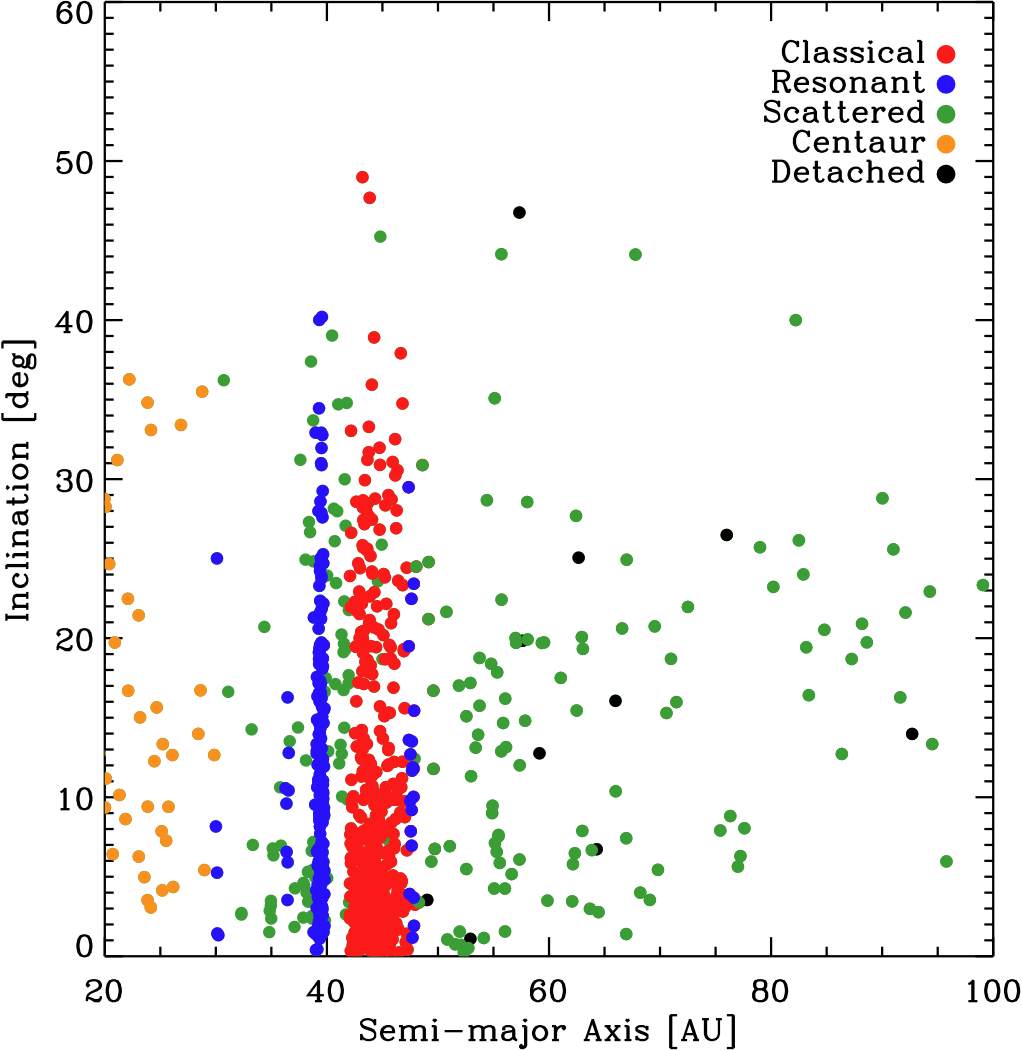}
\caption{Orbital semimajor axis vs. inclination for KBOs known as of 2008
January 30.   The plot shows that the KBOs occupy orbits having a wide range of
inclinations.  Most of the plotted objects were discovered in surveys directed
towards the ecliptic, meaning that an observation bias $\it{against}$ finding
high inclination bodies is imprinted on the sample. }

\label{plotai}       
\end{center}
\end{figure}

The Kuiper belt is a thick disk (Figure \ref{plotai}), with an apparent full
width at half maximum, $FWHM \sim$ 10$^{\circ}$ (Jewitt et al. 1996).  The
apparent width is an underestimate of the true width, however, because most
KBOs have been discovered in surveys aimed near the ecliptic, and the
sensitivity of such surveys varies inversely with the KBO inclination.
Estimates of the unbiased (i.e. true) inclination distribution give $FWHM \sim$
25 to 30$^{\circ}$ (Jewitt et al. 1996, Brown and Trujillo 2001, Elliot et al.
2005 - see especially Figure 20b).  Moreover, all four components of the Kuiper
belt possess broad inclination distributions (Figure \ref{plotai}).  The
inclination distribution of the Classical KBOs appears to be bimodal, with a
narrow core superimposed on a broad halo (Brown and Trujillo 2001, Elliot et
al. 2005). \\

The Kuiper belt is not thin like the Sun's original accretion disk and it is
clear the the inclinations of the orbits of its members have been excited.  The
velocity dispersion amongst KBOs is $\Delta V$ = 1.7 km s$^{-1}$ (Jewitt et al.
1996, Trujillo et al. 2001).   At these velocities, impacts between all but the
largest KBOs lead to shattering and the production of dust, rather than to
accretion and growth.

\begin{svgraybox}

\textbf{Next generation} facilities should provide unprecedented survey capabilities
(through \textit{LSST}) that will provide a deep, all-sky survey of the Solar system to
24th visual magnitude, or deeper.  The number of objects for which reliable
orbits exist will increase from $\sim$10$^3$, at present, by one to two orders
of magnitude, depending on the detailed strategies employed by the surveys.
Large samples are needed to assess the relative populations of the resonances
and other dynamical niches that may place limits on the formation and evolution
of the OSS.  Objects much larger than Pluto, perhaps in the Mars or Earth
class, may also be revealed by careful work.

\end{svgraybox}

\section{Interrelation of the Populations}

Evidence that the small-body populations are interrelated is provided by
dynamical simulations.  The interrelations are of two basic types: a) those
that occur through dynamical processes operating in the current Solar system
and b) those that might have operated at an earlier epoch when the architecture
of the Solar system may have been different from now.  

As examples of the first kind, it is clear that objects in the Oort cloud and
Kuiper belt reservoirs can be perturbed into planet crossing orbits, and that
these perturbations drive a cascade from the outer Solar system through the
Centaurs (bodies, asteroidal or cometary in nature) that are strongly
interacting with the giant planets to the Jupiter family comets (orbits small
enough for the Sun to initiate sublimation of near-surface ice and strongly
interacting with Jupiter) to dead and dormant comets in the near-Earth
``asteroid'' population.   In the current system, simulations show that it is
\textit{not} possible to capture the Trojans or the irregular satellites
(Figure \ref{phoebe}) of the giant planets from the passing armada of small
bodies, and there is no known dynamical path linking comets from the Kuiper
belt, for example, to comets in the main-belt (the MBCs).

\begin{figure}[]
\begin{center}
\includegraphics[width=\textwidth]{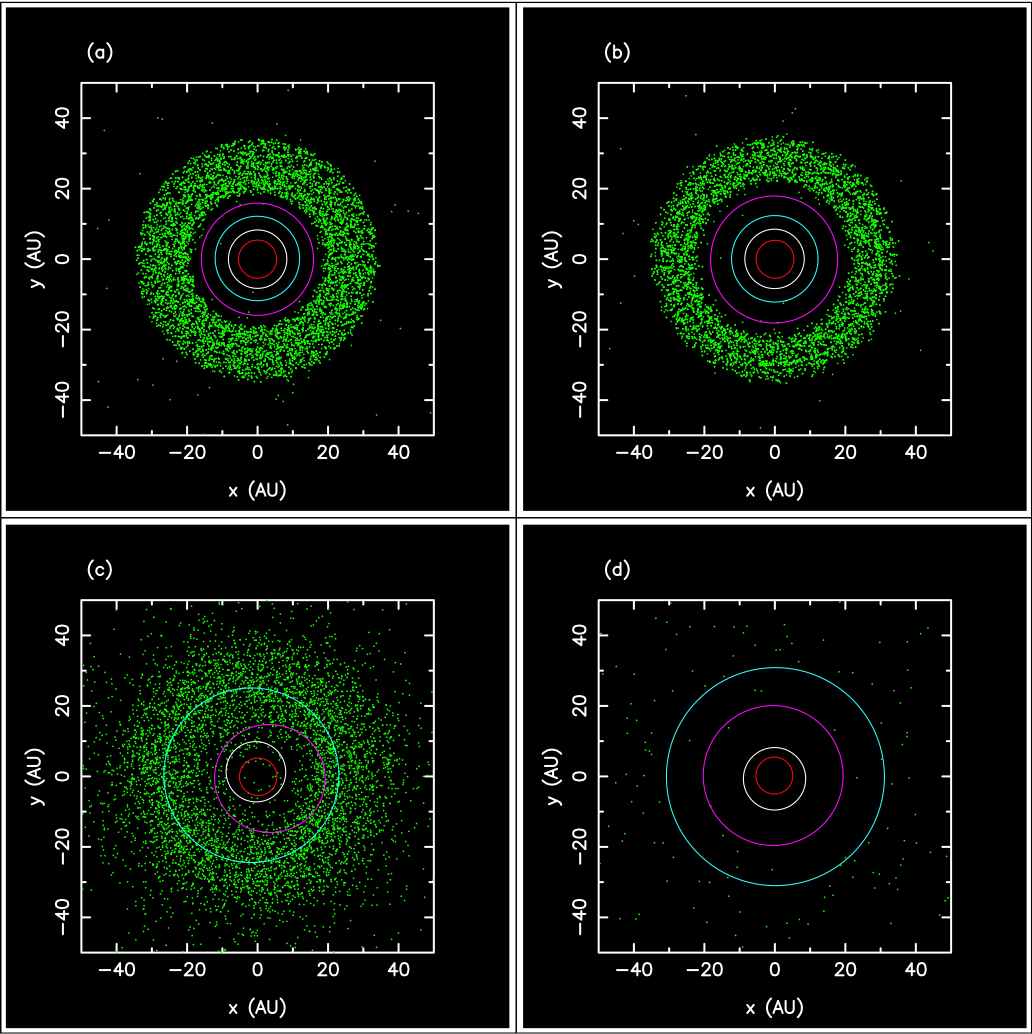}
\caption{``Nice'' model in which the architecture of the Solar system is set by
the clearing of a massive (30 M$_{\oplus}$) Kuiper belt (stippled green region)
when planets are thrown outwards by strong interactions between Jupiter (red)
and Saturn (pink) at the 2:1 mean-motion resonance. (a) The initial
configuration with the giant planets at 5.5, 8.2, 11.5 and 14.2 AU (b) Just
before the 2:1 resonance crossing, timed to occur near 880 Myr from the start
(c) 3 Myr after resonance crossing (note the large eccentricity of Uranus
(purple) at this time and the placement of Neptune (blue) \textit{in} the
Kuiper belt) and (d) 200 Myr later, by which time the planetary orbits have
assumed nearly their current properties.  Adapted with permission from Gomes et
al. 2005}
\label{gomes}       
\end{center}
\end{figure}

The second kind of interrelation is possible because of planetary migration.
The best evidence for the latter is deduced from the resonant Kuiper belt
populations which, in one model, require that Neptune's orbit expanded by
roughly 10 AU, so pushing its mean-motion resonances outwards through the
undisturbed Kuiper belt (Malhotra 1995).  The torques driving the migration
moved Jupiter inwards (by a few $\times$0.1 AU, since it is so massive) as the
other giants migrated outwards.  In one exciting model, the ``Nice model'',
this migration pushed Jupiter and Saturn across the 2:1 mean motion resonance
(Gomes et al. 2005).  The dynamical consequences of the periodic perturbations
induced between the Solar system's two most massive planets would have been
severe (Figure \ref{gomes}).  In published models, these perturbations excite a
30 M$_{\oplus}$ Kuiper belt, placing large numbers of objects into planet
crossing orbits and clearing the Kuiper belt down to its current, puny mass of
$\sim$0.1 M$_{\oplus}$.  During this clearing phase, numerous opportunities
exist for trapping scattered KBOs in dynamically surprising locations.  For
example, the Trojans of the planets could have been trapped during this phase
(Morbidelli et al. 2005).  Some of the irregular satellites might likewise have
been acquired at this time (Nesvorny et al. 2007; furthermore, any irregular
satellites possessed by the planets \textit{before} the mean-motion resonance
crossing would have been lost).  Other KBOs might have been trapped in the
outer regions of the main-asteroid belt, perhaps providing a Kuiper belt source
for ice in the MBCs.  

\begin{figure}[]
\begin{center}
\includegraphics[width=\textwidth]{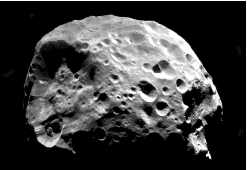}
\caption{Saturn's irregular satellite Phoebe, which might be a captured Kuiper
belt object.  The effective spherical radius is 107$\pm$1 km: the largest
crater, Jason (at left), has a diameter comparable to the radius.  The surface
contains many ices (water, carbon dioxide) and yet is dark, with geometric
albedo 0.08, as a consequence of dust mixed in the ice. Image from Cassini
Imaging Team/NASA/JPL/Space Science Institute. }

\label{phoebe}       
\end{center}
\end{figure}

Whether or not Jupiter and Saturn actually crossed the 2:1 resonance is
unknown.  In the Nice model, the 2:1 resonance crossing is contrived to occur
at about 3.8 Gyr, so as to coincide with the epoch of the late-heavy
bombardment (LHB).  The latter is a period of heavy cratering on the Moon,
thought by some to result from a sudden shower of impactors some 0.8 Gyr after
the formation epoch.   However, the interpretation of the crater age data is
non-unique and reasonable arguments exist to interpret the LHB in other ways
(Chapman et al.  2007).  For example, the cratering rate could merely
\textit{appear} to peak at 3.8 Gyr because all earlier (older) surfaces were
destroyed by an impact flux even stronger than that at the LHB.  (The LHB would
then be an analog to the epoch in the expanding early universe when the optical
depth first fell below unity and the galaxies became visible).

\section{Evidence concerning the birth environment}

Several lines of evidence suggest that the Sun was formed in a dense cluster.  

The existence of widespread evidence for the decay products of short-lived
isotopes in mineral inclusions in meteorites suggests that one or more
supernovae exploded in the vicinity of the Sun, shortly before its formation.
$^{26}$Al (half-life 0.7 Myr) was the first such unstable isotope to be
identified (Lee et al. 1977) but others, like $^{60}$Fe (half-life 1.5 Myr) are
known (and, unlike $^{26}$Al, cannot be produced by nuclear spallation
reactions; Mostefaoui et al. 2004).  It is possible, although not required,
that the collapse formation of the protoplanetary nebula was triggered by shock
compression from the supernova that supplied the unstable nuclei (Ouellette et
al. 2007).

The sharp outer edge to the classical Kuiper belt could be produced by a
stellar encounter having an impact parameter $\sim$150 AU to 200 AU (Ida et al.
2000), although this is only one of several possible causes.  Such close
encounters are highly improbable in the modern epoch but would be more likely
in the dense environment of a young cluster.  At the same time, the survival
and regularity of the orbits of the planets suggests that no very close
encounter occurred (Gaidos 1995).

As noted above, the existence of a dense inner Oort cloud, required in some
models to explain the source of the Halley family comets (long-period comets
which include retrograde examples but which are not isotropically distributed),
can only be populated via the stronger mean perturbations exerted between stars
in a dense cluster.  

Adams and Laughlin (2001) conclude from these and related considerations that
the Sun formed in a cluster of 2000$\pm$1000 stars.  

\section{Colors and Physical Properties}

It has long been recognized that Kuiper belt objects exhibit a diversity in
surface colors unparalleled among Solar system populations (Luu and Jewitt
1996).  The distribution is relatively smooth from neutrally colored to
extremely red objects.  Indeed, a large fraction of KBOs (and Centaurs) is
covered in ``ultrared matter'' (Jewitt 2002), the reddest material observed on
small bodies. This material is absent in other populations, and is thought to
be due to irradiated complex organics (Cruikshank et al. 2007). Further, the
$UBVRIJ$ colors are mutually correlated (Jewitt and Luu 1998; Jewitt et al.
2007), which seems to indicate that the spread is caused by a single reddening
agent. 

The wide range of colors suggests a broad range of surface compositions.
However, such extreme non-uniformity in composition is unlikely to be intrinsic
given the uniform and low temperatures across the disk of the Kuiper belt; the
compositional spread is probably the result of some evolutionary process.
Early theories to explain the color scatter invoke a competition between (the
reddening) long-term exposure to cosmic radiation and (the de-reddening) impact
resurfacing (Luu and Jewitt 1996; Delsanti et al. 2004), but the implied
rotational color variability and correlation between color and collision
likelihood are not observed (Jewitt and Luu 2001; Th{\'e}bault and
Doressoundiram 2003).  

More recently it has been suggested that the diversity was emplaced when the
small body populations were scattered by the outward migration of the ice giant
planets (Gomes 2003), as a consequence of the mutual 2:1 resonance crossing by
Jupiter and Saturn (see Nice model above).  In other words, some of the objects
now in the Kuiper belt may have formed much closer to the Sun, in the 10--20 AU
region, where the chemistry would have been different and perhaps more diverse.
Although appealing, the theory that the KB region was sprinkled with bodies
from various heliocentric distances remains a non-unique explanation and the
implicit assumption that KBOs formed closer to the Sun should be less red
remains ad-hoc.  The relative importances of this dynamical mixing and the
reddening effect by cosmic irradiation are still poorly understood.  Cosmic-ray
reddening seems to be important as the Classical KBOs, supposedly formed
locally at 40 AU and having passively evolving surfaces only subject to cosmic
radiation (their circular orbits protect them from mutual collisions), are on
average the reddest KBOs (Tegler and Romanishin 2000).

A powerful way to investigate the physical properties of KBOs is by the
analysis of their rotational properties, usually inferred from their
lightcurves (Sheppard and Jewitt 2002; Lacerda and Luu 2006). Lightcurves are
periodic brightness variations due to rotation: as a non-spherical (and
non-azimuthally symmetric) KBO rotates in space, its sky-projected
cross-section will vary periodically, and thus modulate the amount of sunlight
reflected back to the observer. The period $P$ and range $\Delta m$ of a KBO's
lightcurve provide information on its rotation period and shape, respectively.
Multi-wavelength lightcurves may also reveal surface features such as albedo or
color patchiness (Buie et al.  1992; Lacerda et al. 2008). These features are
usually seen as second order effects superimposed on the principal,
shape-regulated lightcurve.  By combining the period and the range of a KBO
lightcurve it is possible to constrain its density under the assumption that
the object's shape is mainly controlled by its self-gravity (Jewitt and
Sheppard 2002; Lacerda and Jewitt 2007).  Lightcurves can also uncover
unresolved, close binary objects (Sheppard and Jewitt 2004; Lacerda and Jewitt
2007).

Interesting KBOs, whose lightcurves have been particularly informative include:
1) 134340 Pluto, whose albedo-controlled light variations have been used to map
the distribution of ices of different albedos across the surface (Buie et al.
1992; Young et al. 1999), which is likely controlled by the surface deposition
of frosts from Pluto's thin atmosphere (Trafton 1989), 2) 20000 Varuna, whose
rapid rotation ($P=6.34\pm0.01$ hr) and elongated shape ($\Delta m=0.42\pm0.03$
mag) indicate a bulk density $\rho\sim1000$ kg m$^{-3}$ and hence require an
internally porous structure (Jewitt and Sheppard 2002), 3) 139775 2001 QG298,
whose extreme lightcurve (large range $\Delta m=1.14\pm0.04$ mag and slow
period $P\sim13.8$ hr) suggests an extreme interpretation as a contact or
near-contact binary (Sheppard and Jewitt 2004; Takahashi and Ip 2004; Lacerda
and Jewitt 2007), and finally 4) 136198 2003 EL61, exhibiting super-fast
rotation ($P=3.9$ hr) that requires a density $\sim2500$ kg m$^{-3}$, and a
recently identified surface feature both redder and darker than the average
surface (Lacerda et al. 2008).

Statistically, the rotational properties of KBOs can be used to constrain the
distributions of spin periods (Lacerda and Luu 2006) and shapes (Lacerda and
Luu 2003), which in turn can be used to infer the importance of collisions in
their evolution. For instance, most main-belt asteroids have been significantly
affected by mutual collisions, as shown by their quasi-Maxwellian spin rate
distribution (Harris 1979; Farinella et al.  1981) and a shape distribution
consistent with fragmentation experiments carried out in the laboratory
(Catullo et al. 1984).  KBOs spin on average more slowly ($\langle
P_{KBO}\rangle\sim8.4$ hr vs.  $\langle P_{ast}\rangle\sim6.0$ hr; Lacerda and
Luu 2006) and are more spherical (as derived from the $\Delta m$ distribution;
Luu and Lacerda 2003; Sheppard et al. 2008) than asteroids of the same size,
both indicative of a milder collisional history.  Typical impact speeds in the
current Kuiper belt and main asteroid belt are respectively 2 and 5 km
s$^{-1}$. However, three of the four KBOs listed in the previous paragraph show
extreme rotations and shapes, likely the result of collision events.  Because
the current number density of KBOs is too low for these events to occur on
relevant timescales, their rotations were probably aquired at an early epoch
when the Kuiper belt was more massive and collisions were more frequent (Davis
and Farinella 1997; Jewitt and Sheppard 2002).

\begin{svgraybox}

\textbf{Next generation} telescopes will provide revolutionary new data on the
physical properties of KBOs.  Their colors and rotational properties can 
potentially reveal much about
these objects' surface and physical natures. \textit{JWST}, in particular, will
provide high quality near-infrared spectra that will place the best constraints
on the (probably) organic mantles of the KBOs, thought to be some of the most
primitive matter in the Solar system.  Separate measurements of the albedos and
diameters, obtained from optical/thermal measurements using \textit{ALMA} and \textit{JWST}, will
give the albedos and accurate diameters, needed to fully understand the surface
materials.

Survey data will permit the identification of $>$100 wide binaries, while the
high angular resolution afforded by \textit{JWST} will reveal a much larger number
(thousands?) of close binaries.  Orbital elements for each will lead to the
computation of system masses through Kepler's law.  Diameters from
optical/thermal measurements (using \textit{ALMA} and \textit{JWST}) will then permit the
determination of system densities for KBOs over a wide range of diameters and
orbital characteristics in the Kuiper belt.   Density, as the ``first
geophysical parameter'', provides our best handle on accretion models of the
Kuiper belt objects.  

\end{svgraybox}

\section{Solar System Dust}

Collisions between solids in the early Solar system generally resulted in
agglomeration due to the prevailing low impact energies. However, at present
times, high impact energies dominate and collisions result in fragmentation and
the generation of dust from asteroids, comets and KBOs. Due to the effect of
radiation forces (see $\S$ 7.2) these dust particles spread throughout the
Solar system forming a dust disk. 

\subsection{Inner Solar System: Asteroidal and Cometary Dust}

The existence of dust in the inner Solar system (a.k.a. Zodiacal cloud) has
long been known since the first scientific observations of the Zodiacal light
by Cassini in 1683, correctly interpreted by de Duiliers in 1684 as produced by
sunlight reflected from small particles orbiting the Sun. Other dust-related
phenomena that can be observed naked eye are dust cometary tails and ``shooting
stars". 

The sources of dust in the inner Solar system are the asteroids, as evident
from the observation of dust bands associated with the recent formation of
asteroidal families, and comets, as evident from the presence of dust trails
and tails. Their relative contribution can be studied from the He content of
collected interplanetary dust particles, which is strongly dependent on the
velocity of atmospheric entry, expected to be low for asteroidal dust and high
for cometary dust. Their present contributions are thought to differ by less
than a factor of 10 (Brownlee et al. 1994), but they have likely changed with
time. It is thought that due to the depletion of asteroids, the asteroidal dust
surface area has slowly declined by a factor of 10 (Grogan et al. 2001) with
excursions in the dust production rate by up to an order of magnitude
associated with breakup events like those giving rise to the Hirayama asteroid
families that resulted in the formation of the dust bands observed by {\it
IRAS} (Sykes and Greenberg 1986). The formation of the Veritas family 8.3 Myr
ago still accounts for $\sim$25\% of the Zodiacal thermal emission today
(Dermott et al. 2002). A major peak of dust production in the inner Solar
system is expected to have occurred at the time of the LHB ($\S$4), as a
consequence of an increased rate of asteroidal collisions and to the collisions
of numerous impactors originating in the main asteroid belt (Strom et al. 2005)
with the terrestrial planets. 

The thermal emission of the Zodiacal cloud dominates the night sky between
5--500 $\mu$m and has a fractional luminosity of L$_{dust}$/L$_{Sun}$ $\sim$
10$^{-8}$--10$^{-7}$ (Dermott et al. 2002). Studied by {\it IRAS}, {\it COBE}
and ${\it ISO}$ space telescopes, it shows  a featureless spectrum produced by
a dominant population of low albedo ($<$0.08) rapidly-rotating amorphous
forsterite/olivine grains that are 10--100 $\mu$m in size and are located near
1 AU. The presence a weak (6\% over the continuum) 10 $\mu$m silicate emission
feature indicates the presence of a small population of $\sim$1 $\mu$m grains
of dirty crystalline olivine and hydrous silicate composition (Reach et al.
2003). 

Interplanetary dust particles (IDPs) have been best characterized at around 1
AU by in situ satellite measurements, observations of micro-meteorite impact
craters on lunar samples, ground radar observations of the ionized trails
created as the particles pass through the atmosphere and laboratory analysis of
dust particles collected from the Earth's stratosphere, polar ice and deep sea
sediments. Laboratory analysis of collected IDPs show that the particles are
1--1000$\mu$m in size, typically black, porous ($\sim$40\%) and composed of
mineral assemblages of a large number of sub-micron-size grains with chondritic
composition and bulk densities of 1--3 g/cm$^3$. Their individual origin,
whether asteroidal or cometary, is difficult to establish. The cumulative mass
distribution of the particles at 1 AU follows a broken power-law such that the
dominant contribution to the cross sectional area (and therefore to the
zodiacal emission) comes from 10$^{-10}$ kg grains ($\sim$30 $\mu$m in radius),
while the dominant contribution to the total dust mass comes from $\sim$
10$^{-8}$ kg grains (Leinert and Gr\"un 1990). 

In situ spacecraft detections of Zodiacal dust out to 3 AU, carried out by
Pioneer 8--11, Helios, Galileo and Ulysses, showed that the particles typically
have $i < $30$^{\circ}$ and $e >$ 0.6 with a spatial density falling as
$r^{-1.3}$ for $r <$ 1 AU and $r^{-1.5}$ for $r >$ 1 AU, and that there is a
population of grains on hyperbolic orbits (a.k.a $\beta$-meteoroids), as well
as stream of small grains origina jovian system (see review by Gr\"un et al.
2001).

\subsection{Outer Solar System: Kuiper Belt Dust}

As remarked above, collisions in the modern-day Kuiper belt are erosive, not
agglomerative, and result in the production of dust.  In fact, two components
to Kuiper belt dust production are expected: (1) erosion of KBO surfaces by the
flux of interstellar meteoroids (e.g. Gr\"un et al. 1994), leading to the
steady production of dust at about 10$^3$ to 10$^4$ kg s$^{-1}$ (Yamamoto and
Mukai 1998); and (2) mutual collisions between KBOs, with estimated dust
production of about (0.01--3)$\times$10$^{8}$ kg s$^{-1}$ (Stern 1996). For
comparison, the dust production rate in the Zodiacal cloud, from comets and
asteroids combined, is about 10$^3$ kg s$^{-1}$ (Leinert et al. 1983).  The
fractional luminosity of the KB dust is expected to be around
L$_{dust}$/L$_{Sun}$ $\sim$ 10$^{-7}$--10$^{-6}$ (Stern 1996), compared to
L$_{dust}$/L$_{Sun}$ $\sim$ 10$^{-8}$--10$^{-7}$ for the Zodiacal cloud
(Dermott et al. 2002).

Observationally, detection of Kuiper belt dust at optical wavelengths is
confounded by the foreground presence of cometary and asteroidal dust in the
Zodiacal cloud.  Thermally, these near and far dust populations might be
distinguished on the basis of their different temperatures ($\sim$200 K in the
Zodical cloud vs. $\sim$40 K in the Kuiper belt) but, although sought, Kuiper
belt dust has not been detected this way (Backman et al. 1995).  At infrared
thermal wavelengths both foreground Zodiacal cloud dust \textit{and} background
galactic dust contaminate any possible emission from Kuiper belt dust.  The
cosmic microwave background radiation provides a very uniform source against
which emission from the Kuiper belt might potentially be detected but, again,
no detection has been reported (Babich et al. 2007).

Whereas remote detections have yet to be achieved, the circumstances for
in-situ detection are much more favorable (Gurnett et al. 1997).   The Voyager
1 and 2 plasma wave instruments detected dust particles via the pulses of
plasma created by high velocity impacts with the spacecraft.  Because the
plasma wave detector was not built with impact detection as its primary
purpose, the properties and flux of the impacting dust are known only
approximately.  Still, several important results are available from the Voyager
spacecraft.  Impacts were recorded continuously as the Voyagers crossed the
(then unknown) Kuiper belt region of the Solar system.  The smallest dust
particles capable of generating measurable plasma are thought to be $a_0 \sim$2
$\mu$m in radius.  Measured in the 30 AU to 60 AU region along the Voyager
flight paths, the number density of such particles is $N_1 \sim$
2$\times$10$^{-8}$ m$^{-3}$.  Taking the thickness of the Kuiper belt (measured
perpendicular to the midplane) as $H \sim$ 10 AU, this corresponds to an
optical depth $\tau \sim \pi a_0^2 H N_1 \sim 4\times 10^{-7}$, roughly 10$^3$
times smaller than the optical depth of a $\beta$-Pictoris class dust
circumstellar disk.  An upper limit on the density of gravel ($cm$-sized)
particles in the Kuiper belt is provided by the survival of a 20-cm propellant
tank on the Pioneer 10 spacecraft (Anderson et al. 1998).   

\subsection{Dust Dynamics and Dust Disk Structure}

After the dust particles are released from their parent bodies (asteroids,
comets and KBOs) they experience the effects of radiation and stellar wind
forces. Due to radiation pressure, their orbital elements and specific orbital
energy change  immediately upon release. If their orbital energy becomes
positive ($\beta$ $>$ 0.5), the dust particles escape on hyperbolic orbits
(known as $\beta$-meteoroids -- Zook \& Berg 1975). If their orbital energy
remains negative ($\beta$ $<$ 0.5), their semi-major axis increases but they
remain on bound orbits.  Their new semimajor axis and eccentricity
($\it{a',e'}$) in terms of that of their parent bodies ($\it{a}$ and $\it{e}$)
are $a'=a{1-\beta \over 1-2a\beta/r}$ and $e'={\mid 1 - { (1-2a\beta /r)(1-e^2)
\over (1-\beta^2)}\mid}^{1/2}$ (their inclination does not change), where $r$
is the particle location at release and $\beta$ is the ratio of the radiation
pressure force to the gravitational force. For spherical grains orbiting the
Sun, $\beta=5.7 Q_{pr}/\rho b$, where $\rho$ and $b$ are the density and radius
of the grain in MKS units and $Q_{pr}$ is the radiation pressure coefficient, a
measure of the fractional amount of energy scattered and/or absorbed by the
grain and a function of the physical properties of the grain and the wavelength
of the incoming radiation (Burns, Lamy \& Soter 1979). 

With time, Poynting-Rorbertson (P-R) and solar wind corpuscular drag (which
result from the interaction of the dust grains with the stellar photons and
solar wind particles, respectively) tend to circularize and decrease the
semimajor axis of the orbits, forcing the particles to slowly drift in towards
the central star until they are destroyed by sublimation in a time given by $
t_{PR} = 0.7 ({b \over \mu m}) ({\rho \over kg/m^3}) ({R \over AU})^2 ({L_\odot
\over L_*}) {1 \over 1+albedo} ~yr, $ where $R$ is the starting heliocentric
distance of the dust particle and $\it{b}$ and $\rho$ are the particle radius
and density, respectively (Burns, Lamy and Soter 1979 and Backman and Paresce
1993). If the dust is constantly being produced from a planetesimal belt,  and
because the dust particles inclinations are not affected by radiation forces,
this inward drift creates a dust disk of wide radial extent and uniform
density. Grains can also be destroyed by mutual grain collisions, with a
collisional lifetime of $  t_{col} = 1.26 \times 10^4 ({R \over AU})^{3/2}
({M_\odot \over M_*})^{1/2} ({10^{-5} \over L_{dust}/L_*}) yr $ (Backman and
Paresce 1993).  

For dust disks with M$_{dust}$$>$10$^{-3}$ M$_{\oplus}$, $t_{col} < t_{PR}$,
i.e. the grains are destroyed by multiple mutual collisions before they migrate
far from their parent bodies (in this context, "destruction" means that the
collisions break the grains into smaller and smaller pieces until they are
sufficiently small to be blown away by radiation pressure). This regime is
referred to as collision-dominated. The present Solar system, however,  is
radiation-dominated because it does not contain large quantities of dust and
$t_{col} > t_{PR}$, i.e. the grains can migrate far from the location of their
parent bodies. This is particularly interesting in systems with planets and
outer dust-producing planetesimal belts because in their journey toward the
central star the orbits of the dust particles are affected by gravitational
perturbations with the planets via the trapping of particles in mean motion
resonances (MMRs), the effect of secular resonances and the gravitational
scattering of dust. This results in the formation of structure in the dust disk
(Figure 12). 

\begin{figure}[]
\includegraphics[width=1.0\textwidth]{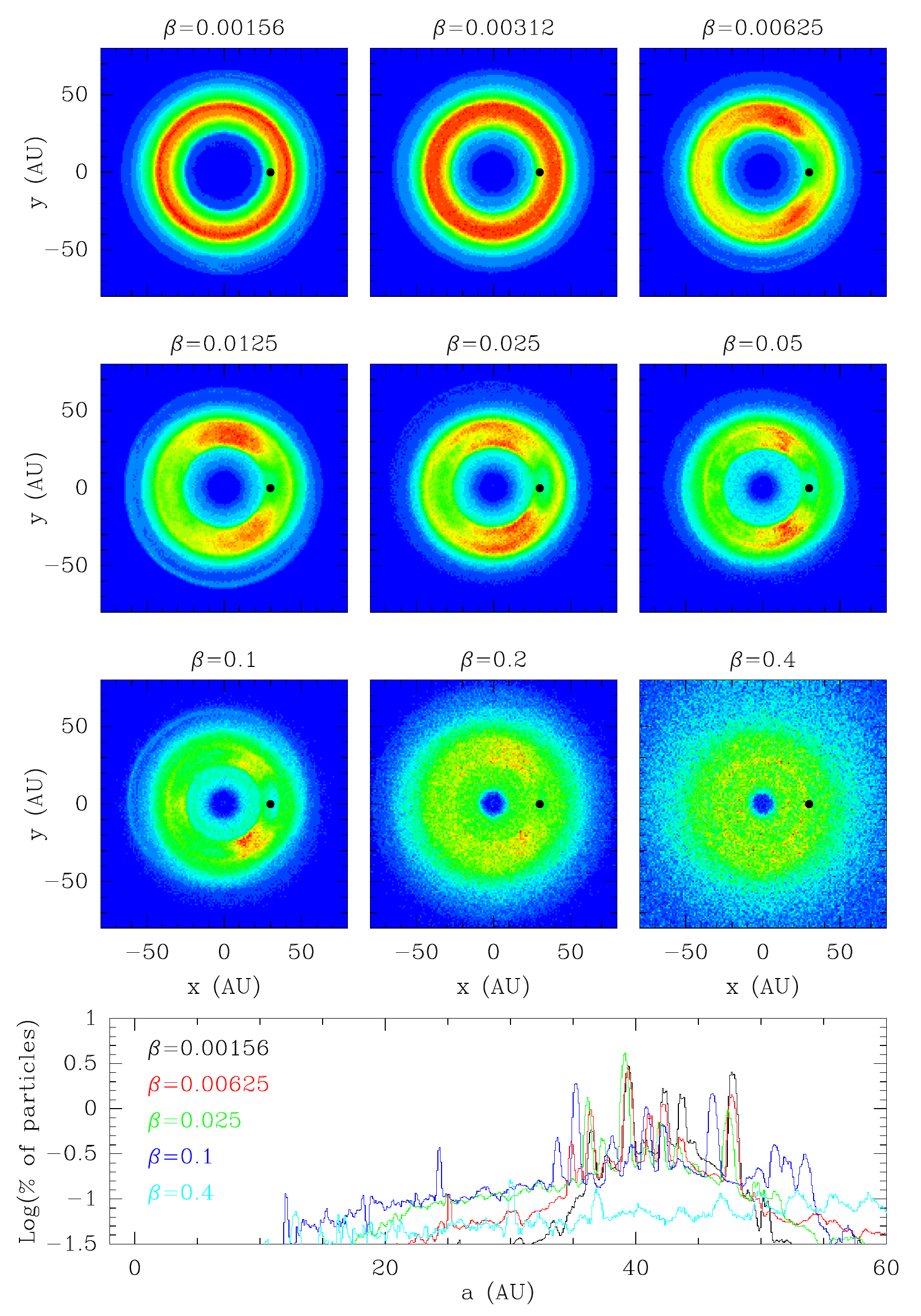}
\caption{Expected number density distribution of the KB dust disk for nine
different particle sizes (or $\beta$ values). $\beta$ is a dimensionless
constant equal to the ratio between the radiation pressure force and the
gravitational force and depends on the density, radius and  optical properties
of the dust grains. If we assume that the grains are composed of spherical
astronomical silicates ($\rho$=2.5, Weingartner \& Draine 2001),  $\beta$
values of 0.4, 0.2, 0.1, 0.05, 0.025, 0.0125, 0.00625, 0.00312, 0.00156
correspond to grain radii of 0.7, 1.3, 2.3, 4.5, 8.8, 17.0, 33.3, 65.9, 134.7
$\mu$m, respectively. The trapping of particles in MMRs with Neptune is
responsible for the ring-like structure, the asymmetric clumps along the orbit
of Neptune, and the clearing of dust at Neptune's location (indicated with a
black dot). The disk structure is more prominent for larger particles (smaller
$\beta$ values) because the P-R drift rate is slower and the trapping is more
efficient. The disk is more extended in the case of small grains (large $\beta$
values) because small particles are more strongly affected by radiation
pressure. The histogram shows the relative occurrence of the different MMRs for
different sized grains, where the large majority of the peaks correspond to
MMRs with Neptune. The inner depleted region inside $\sim$ 10 AU is created by
gravitational scattering of dust grains with Jupiter and Saturn. More details
on these models can be found in Moro-Mart\'{\i}n \& Malhotra (2002, 2003).}
\label{KBdust}       
\end{figure}

Dust particles drifting inward can become entrapped in exterior MMRs because at
these locations the particle receives energy from the perturbing planet that
can balance the energy loss due to P-R drag, halting the migration. This makes
the lifetime of particles trapped in outer MMRs longer than in inner MMRs (Liou
\& Zook 1997), with the former dominating the disk structure. This results in
the formation of resonant rings outside the planet's orbit, as the vast
majority of the particles spend most of their lifetimes trapped in exterior
MMRs. In some cases,  due to the geometry of the resonance,  a clumpy structure
is created. Figure \ref{KBdust} (from Moro-Mart\'{\i}n and Malhotra 2002) shows
the effect of resonant trapping expected in the (yet to be observed) KB dust
disk, where the ring-like structure, the asymmetric clumps along the orbit of
Neptune, and the clearing of dust at Neptune's location are all due to the
trapping of particles in MMRs with Neptune (as seen in the histogram of
semimajor axis). Neptune plays the leading role in the trapping of dust
particles because of its mass and because it is the outermost planet and its
exterior resonances are not affected by the interior resonances from the other
planets. Trapping is more efficient for larger particles (i.e. smaller $\beta$
values) because the drag force is weaker and the particles cross the resonance
at a slower rate increasing their probability of being captured.  The effects
of resonance trapping in hypothetical planetary systems each consisting of a
single planet on a circular orbit and an outer planetesimal belt similar to the
KB are shown in Figure \ref{FIX}. More eccentric planets ($e < 0.6$)
can also create clumpy eccentric rings and offset rings with a pair of clumps
(Kuchner and Holman 2003). Even though, as mentioned in $\S$7.2, the KB dust
disk has yet to be observed, this is not the case for the Zodical cloud, for
which {\it IRAS}  and {\it COBE} thermal observations show that there is a ring
of asteroidal dust particles trapped in exterior resonances with the Earth at
around 1 AU, with a 10\% number density enhancement on the Earth's wake that
results from the resonance geometry (Dermott et al. 1994).  

\begin{figure}[]
\includegraphics[width=1.0\textwidth]{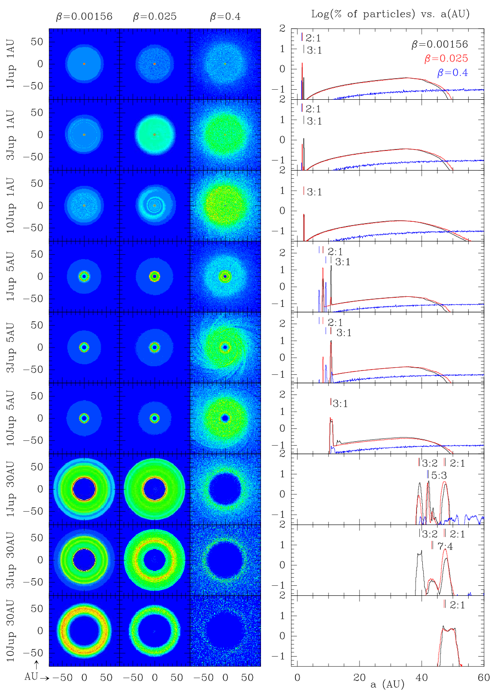}
\caption{Same as Figure 12 but for nine hypothetical planetary systems around a
solar type star consisting of a single planet with a mass of 1, 3 or
10M$_{Jup}$  in a circular orbit at 1, 5 or 30 AU, and a coplanar outer
planetesimal belt similar to the KB. The models with 1 M$_{Jup}$ planet at 1
and 5 AU show that the dust particles are preferentially trapped in the 2:1 and
3:1 resonances, but when the mass of the planet is increased to 10 M$_{Jup}$,
and consequently the hill radius of the planet increases, the 3:1 becomes
dominant. The resonance structure becomes richer when the planet is further
away from the star, and when the mass of the planet decreases (compare the
model for a 1M$_{Jup}$ planet at 30 AU with the Solar system model in Figure
12, where the structure is dominated by Neptune). From Moro-Mart\'{\i}n, Wolf
\& Malhotra (in preparation). 
}
\label{singleplanet}       
\end{figure}

Secular perturbations, the long-term average of the perturbing forces,  act on
timescales $>$0.1 Myr. If the planet and the planetesimal disk are not
coplanar, the secular perturbations can create a warp in the dust disk as their
tendency to align the orbits operates on shorter timescales closer to the star.
A warp can also be created in systems with two non-coplanar planets. If the
planet is in an eccentric orbit,  the secular resonances can force an
eccentricity on the dust particles and this creates an offset in the disk
center with respect to the star that can result in a brightness asymmetry.
Other effects of secular perturbations are spirals and inner gaps (Wyatt et al.
1999). The effect of secular perturbations can be seen in {\it IRAS}  and {\it
COBE} observations on the Zodiacal cloud and account for the presence of an
inner edge around 2 AU due to a secular resonance with Saturn (that also
explains the inner edge of the main asteroid belt), the offset of the cloud
center with respect to the Sun, the inclination of the cloud with respect to
the ecliptic, and the cloud warp (see review in Dermott et al. 2001). 

\begin{figure}[]
\begin{center}
\includegraphics[width=1.0\textwidth]{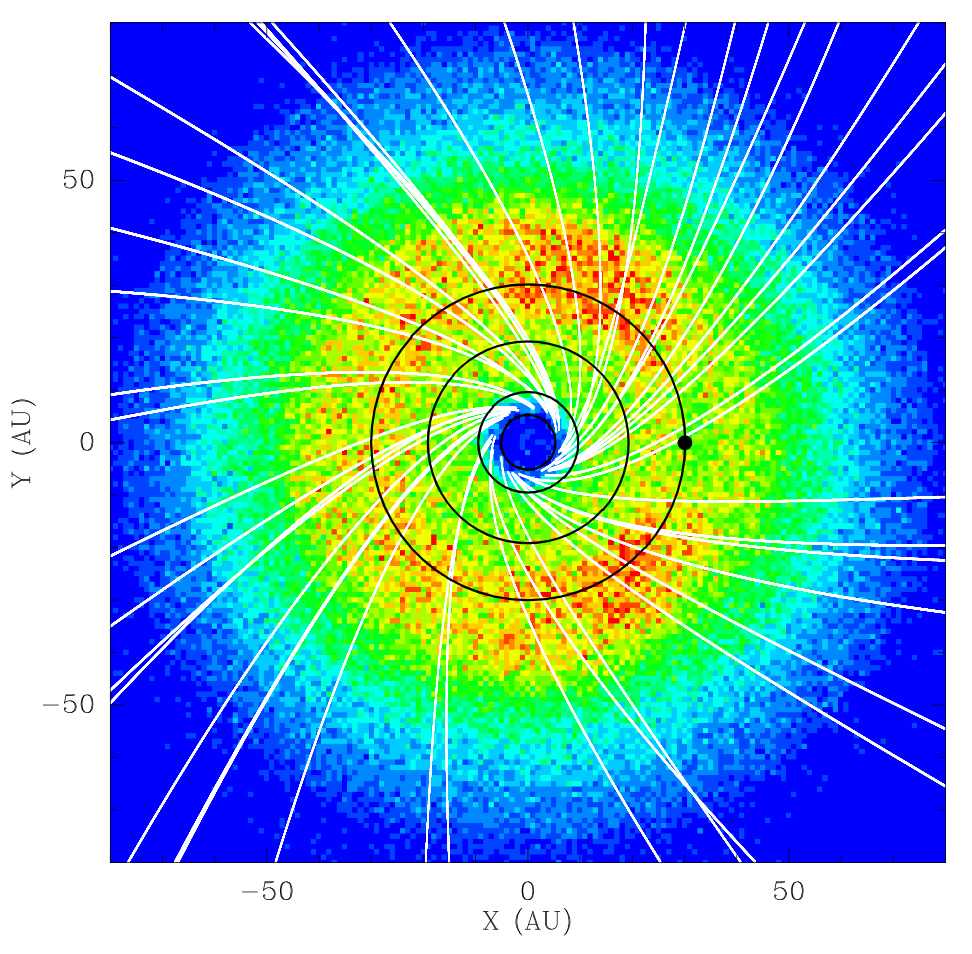}
\caption{Expected number density distribution of a KB dust disk composed of
particles with $\beta$ = 0.2 with the trajectories of the particles ejected by
Jupiter in white. The black dot indicates the position of Neptune and the
circles correspond to the orbits of the Giant planets. In addition to the
population of small grains with $\beta$ $>$ 0.5 blown-out by radiation
pressure, the gravitational scattering by the giant planets (Jupiter and Saturn
in the case of the Solar system) produces an outflow of large grains ($\beta$
$<$ 0.5) that is largely confined to the ecliptic (Moro-Mart\'{\i}n \& Malhotra
2005b).  Interestingly, a stream of dust particles arriving from the direction
of $\beta$ Pictoris has been reported by Baggaley (2000).  }
\label{wind}       
\end{center}
\end{figure}

\begin{figure}[]
\begin{center}
\includegraphics[width=0.95\textwidth]{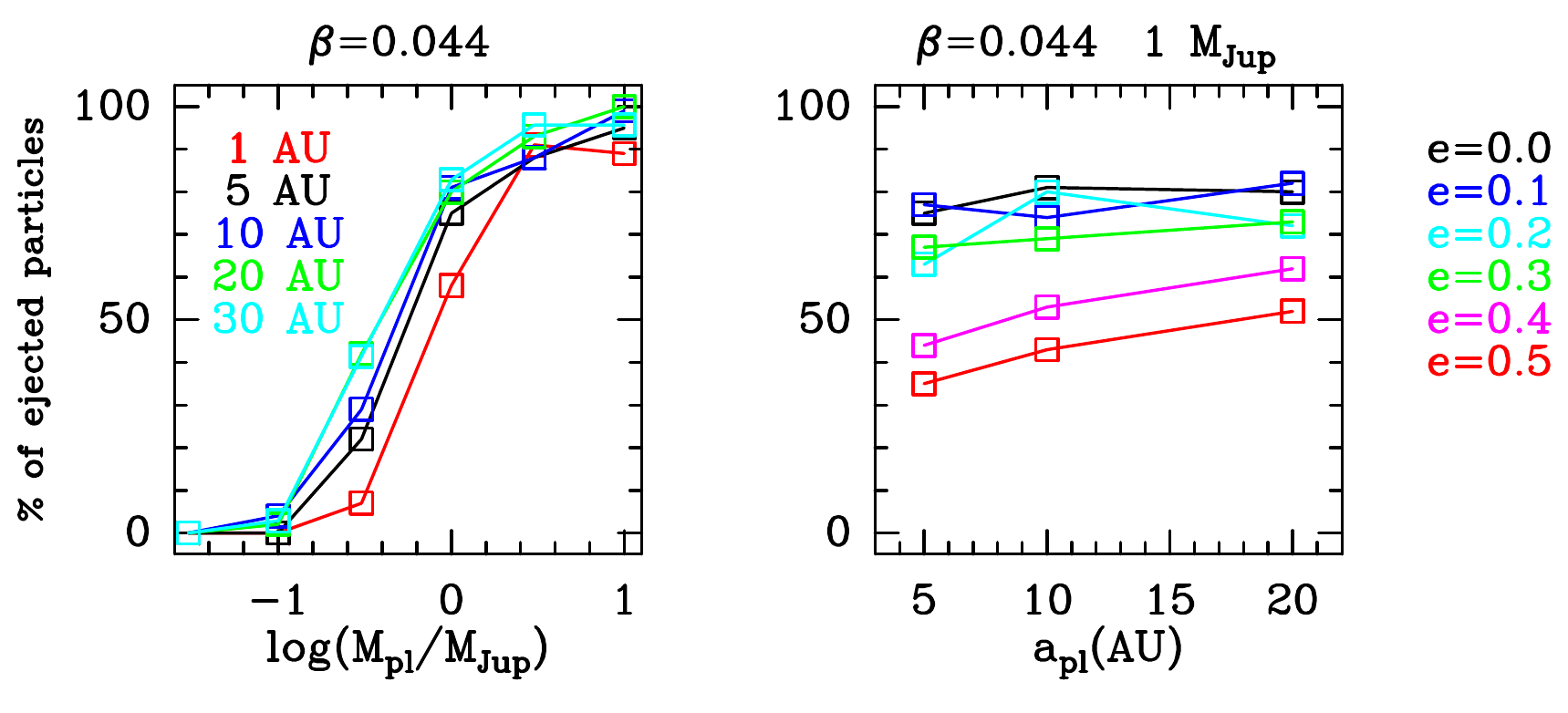}
\caption{Percentage of dust particles ejected from the system by gravitational
scattering with the planet for the single planet models in Figure
\ref{singleplanet}.  The particle size is fixed, corresponding to grains with
$\beta$ = 0.044. Because gravitational scattering is independent of the
particle size, the efficiency of ejection is fairly independent of $\beta$.
{\it Left}: Dependency of the efficiency of ejection on the planet's mass
(x-axis) and the planet's semimajor axis (indicated by the different colors).
{\it Right}: Dependency of the efficiency of ejection on the planet semimajor
axis (x-axis) and eccentricity (corresponding to the different colors). The
models in the right panel correspond to a 1 M$_{Jup}$ mass planet on a circular
orbit around a solar type star. Planets with masses of 3--10 M$_{Jup}$ at 1
AU--30 AU in a circular orbit eject $>$90\% of the dust grains that go past
their orbits under P-R drag; a 1 M$_{Jup}$ planet at 30 AU ejects $>$80\% of
the grains, and about 50\%--90\% if located at 1 AU, while a 0.3 M$_{Jup}$
planet is not able to open a gap, ejecting $<$ 10\% of the grains. These
results are valid for dust grains sizes in the range 0.7~$\mu$m--135~$\mu$m.
From Moro-Mart\'{\i}n and Malhotra 2005.  }
\label{efficiency}       
\end{center}
\end{figure}

The efficient ejection of dust grains by gravitational scattering with massive
planets as the particles drift inward from an outer belt of planetesimals
(Figures \ref{wind} and \ref{efficiency}) can result in the formation of a dust
depleted region inside the orbit of the planet (Figures \ref{KBdust} and
\ref{singleplanet}), such as the one expected inside 10 AU in the KB dust disk
models (due to gravitational scattering by Jupiter and Saturn).  

\begin{svgraybox}

\textbf{Next generation} facilities will probably be unable to detect diffuse emission from Kuiper dust because the optical depth is so low, and the effects of
foreground and background confusion so large.  However, small Kuiper belt
objects are sufficiently numerous that there is a non-negligible chance that the
collision clouds of recent impacts will be detected (Stern 1996).  Such clouds
can potentially be very bright, and evolve on timescales (days and weeks) that
are amenable to direct observational investigation using \textit{LSST} and \textit{JWST}.
Collision cloud measurements provide our best chance to understand the
sub-kilometer population in the Kuiper belt.  These objects are too small to be
directly detected but are of special relevance as the precursors to the
Centaurs and Jupiter family comet nuclei.  Measurement of their number is of
central importance in understanding the role of the Kuiper belt as the JFC
source, and as the source of Kuiper dust.

\end{svgraybox}

\section{Kuiper Belts of Other Stars}

Radial velocity studies have revealed that $>$7\% of solar-type stars harbor
giant planets with masses $<$13 M$_{Jup}$ and semimajor axis $<$ 5 AU (Marcy et
al. 2005).  This is a lower limit because the duration of the surveys (6--8
years) limits the ability to detect planets long-period planets; the expected
frequency extrapolated to 20 AU is $\sim$12\% (Marcy et al. 2005).  As of
February 2008, 276 extra-solar planets have been detected with a mass
distribution that follows d$\it{N}$/d$\it{M}$ $\propto$$\it{M}$$^{-1.05}$ from
0.3M$_{Jup}$ to 10 M$_{Jup}$ (the surveys are incomplete at smaller masses). A
natural question arises whether these planetary systems, some of them harboring
multiple planets, also contain planetesimals like the asteroids, comets and
KBOs in the Solar system.  Long before extra-solar planets were discovered we
inferred that the answer to this question was yes: colliding planetesimals had
to be responsible for the dust disks observed around mature stars. In $\S$8.1
we will discuss how these dust disks, known as {\it debris disks}, can help us
study indirectly Kuiper belts around other stars; other methods by which
extra-solar Kuiper belts might be found and characterized in the future will be
discussed in $\S$8.2 and $\S$8.3.

\subsection{Debris Disks}
\subsubsection{Evidence of Planetesimals}

Theory and observations show that stars form in circumstellar disks composed of
gas and dust that had previously collapsed from the densest regions of
molecular clouds.  For solar type stars, the masses of these disks are $\sim$
0.01--0.10 M$_{\odot}$ and extend to 100s of AU, comparable to the minimum mass
solar nebula ($\sim$0.015 M$_{\odot}$ , which is the mass required to account
for the condensed material in the Solar system planets).  Over time, these
primordial or proto-planetary disks, with dust grain properties similar to
those found in the interstellar medium,  dissipate as the disk material
accretes onto the star, is blown away  by stellar wind ablation,
photo-evaporation or high-energy stellar photons, or is stripped away by
passing stars. The primordial gas and dust in these disks dissipate in less
than 10$^{7}$ years (see e.g. Hartmann 2000 and references therein).

However, it is found that some main sequence stars older than $\sim$10$^{7}$
years show evidence of dust emission. In most cases, this evidence comes from
the detection of an infrared flux in excess of that expected from the stellar
photosphere, thought to arise from the thermal emission of circumstellar dust.
In some nearby stars, like the ones shown in Figures \ref{debrisdisks},
\ref{aumic} and \ref{vega},  direct imaging has confirmed that the emission
comes from a dust disk. In $\S$7.3 we discussed the lifetimes of the dust
grains due to radiation pressure, P-R drag and mutual grain collisions. It is
found that in most cases these lifetimes are much shorter than the age of the
star\footnote{One needs to be cautious with this argument because the ages of
main sequence stars are difficult to determine. For example, the prototype (and
best studied) of debris disk $\beta$-Pictoris, was dated at $\sim$100 to 200
Myr (Paresce 1991) but later become only $\sim$20 Myr old (Barrado y
Navascu{\'e}s et al. 1999).}, and therefore the observed dust cannot be
primordial but is more likely produced by a reservoir of undetected
dust-producing planetesimals, like the KBOs, asteroids and comets in the Solar
system (see e.g. Backman and Paresce 1993). This is why these dust disks
observed around mature main sequence stars are known as debris disks. Debris
disks are evidence of the presence of planetesimals around other main sequence
stars. In the core accretion model, these planetesimals formed in the earlier
protoplanetary disk phase described above, as the ISM-like dust grains
sedimented into the mid-plane of the disk and aggregated into larger and larger
bodies (perhaps helped by turbulence) until they became planetesimals, the
largest of which could potentially become the seeds out of which the giant
planets form from the accretion of gas onto these planetary cores. 

Even though these extra-solar planetesimals remain undetected, the dust they
produce has a much larger cumulative surface area that makes the dust
detectable in scattered light and in thermal emission. The study of these
debris disks can help us learn indirectly about their parent planetesimals,
roughly characterizing their frequencies, location and composition, and even
the presence of massive planets.

\subsubsection{Spatially Resolved Observations}

\begin{figure}[]
\begin{center}
\includegraphics[width=0.9\textwidth]{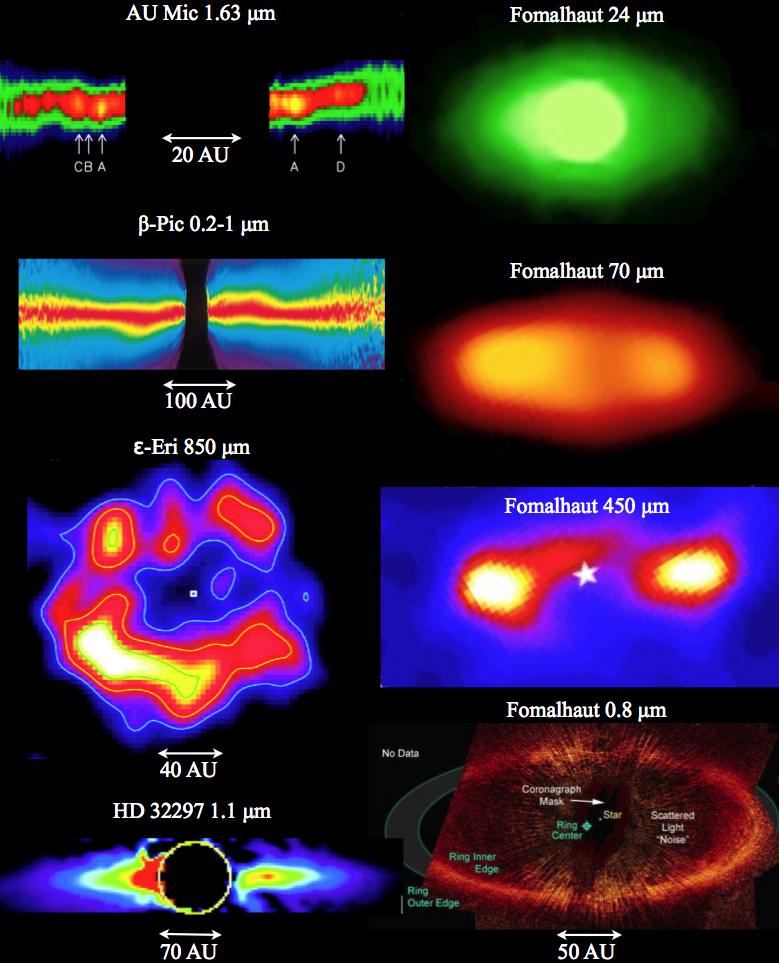}
\caption{Spatially resolved images of nearby debris disks showing dust emission
from 10s to 100s of AU with a wide diversity of complex features including
inner gaps, warps, brightness asymmetries, offsets and clumply rings, some of
which may be due to the presence of massive planets ($\S$7.3). {\it Left} (from
top to bottom): AU-Mic (Keck AO at 1.63 ~$\mu$m; Liu 2004), $\beta$-Pic (STIS
CCD coronography at 0.2--1~$\mu$m; Heap et al. 2000), $\epsilon$-Eri
(JCMT/SCUBA at 850~$\mu$m; Greaves et al. 2005) and HD 32297 (HST/NICMOS
coronography at 1.1~$\mu$m; Schneider, Silverstone and Hines 2005). {\it
Right}. All images correspond to Fomalhaut (7.7 pc away) and are on the same
scale. From top to bottom: Spitzer/MIPS at 24~$\mu$m (Stapelfeldt et al. 2004);
Spitzer/MIPS at 70~$\mu$m (Stapelfeldt et al. 2004); JCMT/SCUBA at 450~$\mu$m
(Holland et al. 2003) and HST/ACS at 0.69--0.97~$\mu$m (Kalas et al. 2005). For
this last panel, the annular disk in the scattered light image has an inner
radius of $\sim$133 AU and a radial thickness of $\sim$25 AU and its center is
offset from the star by about 15$\pm$1 AU in the plane, possibly induced by an
unseen planet. Its sharp inner edge has also been interpreted as a signature of
a planet (Kalas et al. 2005).  The dust mass of the Fomalhaut debris disk in
millimeter-sized particles is about 10$^{23}$ kg ($\sim$0.02 M$_{\oplus}$;
Holland et al. 1998) but, if larger bodies are present, the mass could be 50 to
100 M$_{\oplus}$, or about 1000 times the mass of our Kuiper belt.  }
\label{debrisdisks}       
\end{center}
\end{figure}

\begin{figure}[]
\begin{center}
\includegraphics[width=\textwidth]{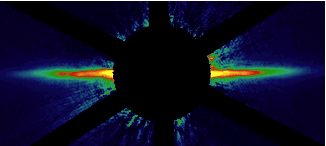}
\caption{HST image of the AU Mic disk with the central regions obscured by a
coronagraphic mask.  AU Mic is a 12$_{-4}^{+8}$ Myr old M dwarf ($\sim$0.5
M$_{\odot}$) only 9.9$\pm$0.1 pc from Earth. Its excess thermal emission at
submillimeter wavelengths suggests a dust mass near 0.01 M$_{\oplus}$ (Kalas et
al. 2004) while, in scattered light, it shows a nearly edge-on disk about 100
AU in diameter with evidence for structure (Liu 2004). From Liu et al. (2004).
}

\label{aumic}       
\end{center}
\end{figure}

\begin{figure}[]
\begin{center}
\includegraphics[width=0.8\textwidth]{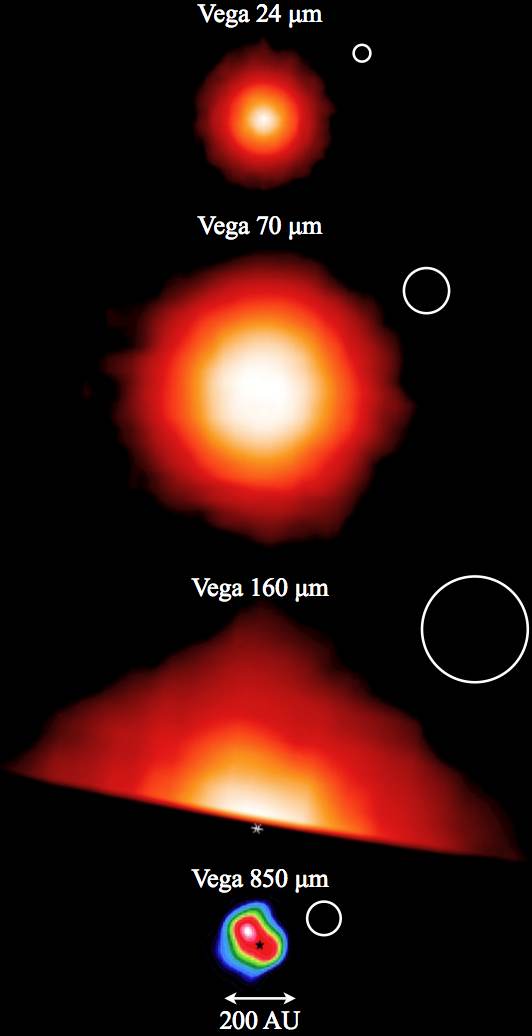}
\caption{Spatially resolved images of Vega from Spitzer/MIPS at 24, 70,
160~$\mu$m (Su et al. 2005) and from JCMT/SCUBA at 850~$\mu$m (Holland et al.
1998). All images are in the same scale. The instrument beam sizes (shown in
white circles) indicate that the wide radial extent of  the MIPS disk images
compared to the SCUBA disk image is not a consequence of the instrumental PSF
but due to a different spatial location of the particles traced by the two
instruments. The sub-mm emission is thought to arise from large dust particles
originating from a planetesimal belt analogous to the KB, while the MIPS
emission is though to correspond to smaller particles with $\beta$$<$0.5 (may
be due to porosity), produced by collisions in the planetesimal belt traced by
the sub-mm observations, and that are blown away by radiation pressure to
distances much larger than the location of the parent bodies (Su et al. 2005).
This scenario would explain not only the wider extent of the MIPS disk but also
its uniform distribution, in contrast with the clumply and more compact sub-mm
disk.  }
\label{vega}       
\end{center}
\end{figure}

Most debris disk observations are spatially unresolved and the debris disks are
identified from the excess thermal emission contributed by dust in their
spectral energy distributions (SEDs). 
In a few cases (about two dozen so far), the disks are close enough and the
images are spatially resolved.  Figures \ref{debrisdisks}, \ref{aumic} and
\ref{vega} show the most spectacular examples.  These high resolution
observations show a rich diversity of morphological features including warps
(Au-Mic \& $\beta$-Pic), offsets of the disk center with respect to the central
star ($\epsilon$-Eri and Fomalhaut), brightness asymmetries (HD 32297 and
Fomalhaut), clumpy rings (Au-Mic, $\beta$-Pic, $\epsilon$-Eri and Fomalhaut)
and sharp inner edges (Fomalhaut), features that, as discussed in $\S$7.3 could
be due to gravitational perturbations of massive planets, and that in some
cases have been observed in the zodiacal cloud, while in other cases used to
belong to the realm of KB disk models. Even though the origin of individual
features is still under discussion and the models require further refinements
(e.g. in the dust collisional processes and the effects of gas drag), the
complexity of these features, in particular the azimuthal asymmetries, indicate
that planets likely play a role in the creation of structure in the debris
disks.  This is of interest because the structure, in particular that created
by the trapping of particles in MMRs, is sensitive to the presence of
moderately massive planets at large distances (recall the KB dust disk models
and the structure created by Neptune). This is a parameter space that cannot be
explored with the present planet detection techniques, like the radial velocity
and transient studies, and therefore the study of debris disk structure can
help us learn about the diversity of planetary systems. In this context, high
resolution debris disk observations over a wide wavelength range are of
critical importance, like those to be obtained with {\it Herschel}, {\it JWST}
and {\it ALMA}. 

\subsubsection{Spectral Energy Distributions}

As we mentioned above, most of the debris disks observations are spatially
unresolved and are limited to the study of the SED of the star+disk system.
Even assuming that the dust is distributed in a disk (and not, for example, in
a spherical shell) there are degeneracies in the SED analysis and the dust
distribution cannot be unambiguously determined (e.g. Moro-Moro-Mart\'{\i}n,
Wolf \& Malhotra 2005a).  Nevertheless, a wealth of information can be
extracted from the SED.  {\it IRAS} and {\it ISO} made critical discoveries on
this front, but the number of known debris disks remained too small for
statistical studies. This changed recently with the unprecedented sensitivity
of the {\it Spitzer} instruments, that allowed the detection of hundreds of
debris disks in large stellar surveys that searched for dust around 328 single
FGK stars (Hillenbrand et al. 2008, Meyer et al.  2008, Carpenter et al. in
preparation), a different sample of 293 FGK  stars (Trilling et al. 2008,
Beichman et al. 2006a, 2006b, Bryden et al. 2006), 160 A single stars (Su et
al. 2006, Rieke et al. 2005); 69 A3--F8 binary stars (Trilling et al. 2007),
and in young stellar clusters (Gorlova et al. 2007, Siegler et al. 2007). As a
result, we now possess information concerning their frequencies, their
dependency with stellar type and stellar environment, their temporal evolution
and the composition of the dust grains (see e.g. Moro-Mart\'{\i}n et al. 2007
for a recent review).\\ 
\\
\\
\\

\noindent{\it Debris Disk Frequencies}\\

\begin{figure}[]
\begin{center}
\includegraphics[width=0.7\textwidth]{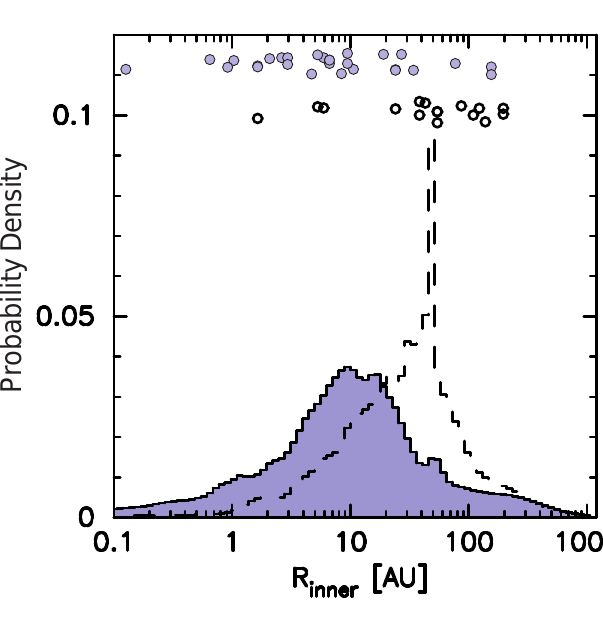}
\caption{Probability distribution for disk inner radii based on the analysis of
the spectra (12--35  $\mu$m) of 44 debris disks around FGK stars from the {\it
FEPS} survey. The dashed and grey histograms correspond to sources with and
without 70 $\mu$m excess, respectively (with best fit parameters are shown as
open and grey circles). Typical disk inner radius  are $\sim$ 40 AU and $\sim$
10 AU for disks with and without 70 $\mu$m excess, respectively, indicating
that most of the debris disks observed are KB-like.  Figure adapted with
permission from Carpenter et al. (in preparation). }
\label{carpenter1}       
\end{center}
\end{figure}

\begin{figure}[]
\begin{center}
\includegraphics[width=0.9\textwidth]{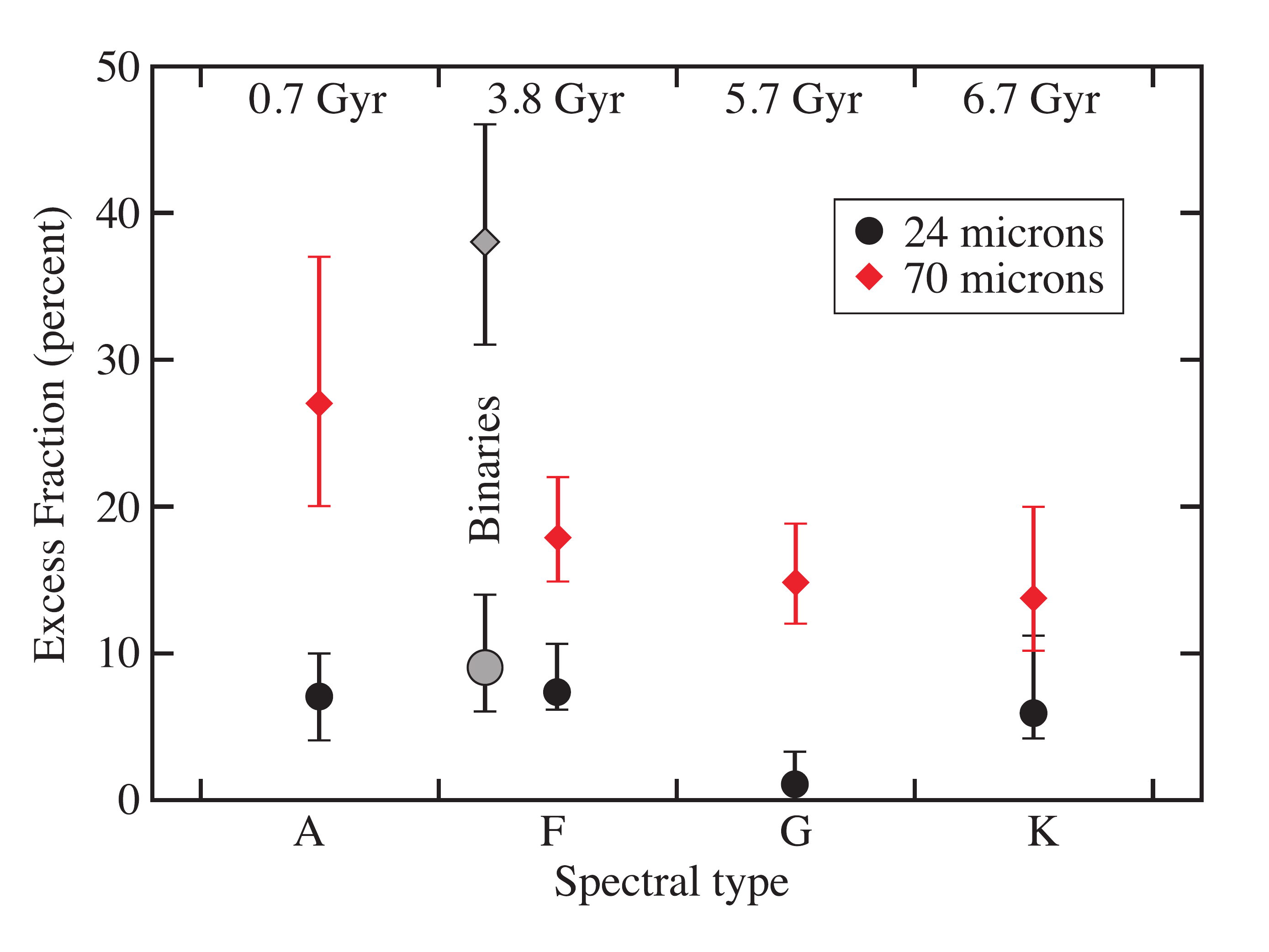}
\caption{The percentage of stars showing excess dust emission, i.e. with
indirect evidence of the presence of dust-producing planetesimal belts, as a
function of stellar type for ages $>$ 600 Myr (the mean ages within each type
are shown at the top). The vertical bars correspond to binomial errors that
include 68\% of the probability (1 $\sigma$ for Gaussian errors). Black is for
24 $\mu$ excess emission (tracing warmer dust) and red is for 70 $\mu$ (tracing
colder dust).  The data are consistent with no dependence on spectral type.
There seems to be a weak decrease with spectral type at 70 $\mu$ but so far
this is statistically not significant and could be due to an effect of age.
Further analysis indicates that percentage of stars showing excess is different
in old A stars and in M stars than in FGK stars. The excess rate for old M
stars is 0\% with upper limits (binomial errors) of 2.9\% at 24 $\mu$m and 12\%
at 70 $\mu$m (Gautier et al. 2007). The lack of distant disks around K stars
may be an observational bias because their peak emission would be at $\lambda$
$>$ 70 $\mu$m and therefore remain undetected by {\it Spitzer}. The upcoming
{\it Herschel} space telescope will provide the sensitivity to explore more
distant and fainter debris disks. Figure adapted 
from Trilling et al. (2008) with data from Su et al.  (2006), Trilling et al.
(2007), Beichman et al.  (2006b) and Gautier et al.  (2007).  }
\label{trilling11}       
\end{center}
\end{figure}

The {\it Spitzer FEPS}\footnote{http://feps.as.arizona.edu/} survey of 328 FGK
stars found that the frequency of 24 $\mu$m excess is 14.7\% for stars younger
than 300 Myr and  2\% for older stars, while at 70 $\mu$m, the excess rates are
6--10\% (Hilllenbrand et al. 2008 and Carpenter et al. in preparation).  These
disks show characteristic temperatures of 60--180 K with evidence of a
population of colder grains to account for the 70 $\mu$m excesses; the implied
disk inner radii are $>$ 10 AU and extend over tens of AU (see Figure
\ref{carpenter1} -- Carpenter et al. in preparation).  Figure \ref{trilling11}
shows the debris disks incidence rates derived from a combined sample of 350
AFGKM stars from Trilling et al. (2008); for the 225 Sun-like (FG) stars in the
sample older than 600 Myr, the frequency of the debris disks are
4.2$^{+2}_{-1.1}$\% at 24 $\mu$m and 16.4$^{+2.8}_{-2..9}$\% at 70 $\mu$m. 

The above debris disks incidence rates compare to $\sim$20\% of solar-type
stars that harbor giant planets inside 20 AU (Marcy et al. 2005).  Even though
the frequencies seem similar, one should keep in mind that the sensitivity of
the {\it Spitzer} observations is limited to fractional luminosities of
L$_{dust}$/L$_*$$>$10$^{-5}$, i.e. $>$100 times the expected luminosity from
the KB dust in our Solar system. Assuming a gaussian distribution of debris
disk luminosities and extrapolating from {\it Spitzer} observations (showing
that the frequency of dust detection increases steeply with decreasing
fractional luminosity), Bryden et al. (2006) found that the luminosity of the
Solar system dust is consistent with being 10 $\times$ brighter or fainter than
an average solar-type star, i.e. debris disks at the Solar system level could
be common. Observations therefore indicate that planetary systems harboring
dust-producing KBOs are more common than those with giant planets, which would
be in agreement with the core accretion models of planet formation where the
planetesimals are the building blocks of planets and the conditions required
for to form planetesimals are less restricted than those to form gas giants.
Indeed, there is no apparent difference between the incidence rate of debris
disks around stars with and without known planetary companions
(Moro-Moro-Mart\'{\i}n et al.  2007,  Bryden et al. in preparation), although
planet-bearing stars tend to harbor more dusty disks (Bryden et al. in
preparation), which could result from the excitation of the planetesimals'
orbits by gravitational perturbations with the planet.  Figure \ref{trilling11}
shows that there is no dependency on stellar type, neither in the frequency of
debris disks, nor on the dust mass and location, indicating that planetesimal
formation can take place under a wide range of conditions (Trilling et al.
2008).  \\

\noindent{\it Debris Disk Evolution}\\

\begin{figure}[]
\begin{center}
\includegraphics[width=0.9\textwidth]{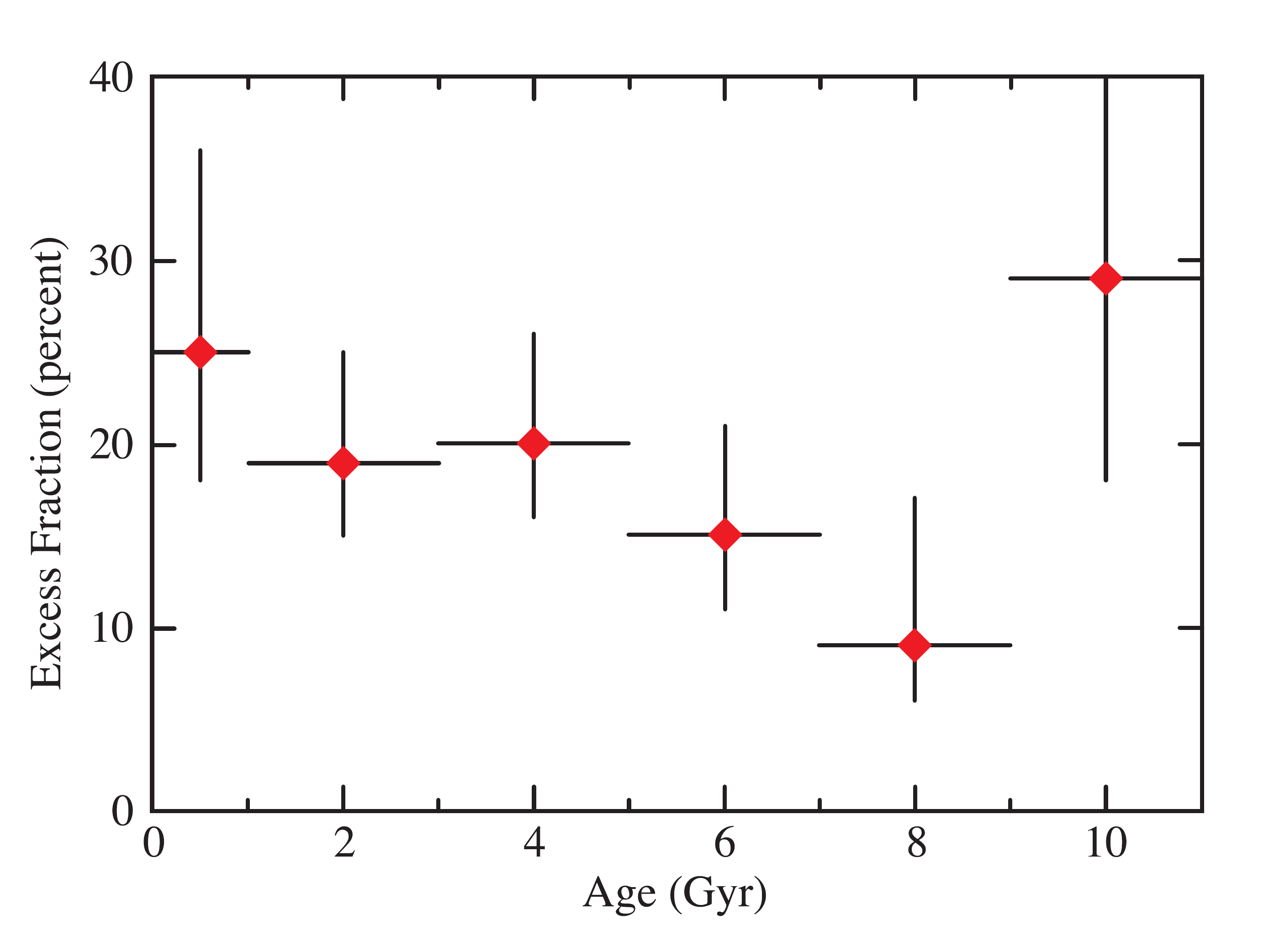}
\caption{The percentage of stars showing excess dust emission, i.e. with
indirect evidence of the presence of dust-producing planetesimal belts, as a
function of age for the F0--K5 stars. The horizontal error bars are the age
bins (not the age uncertainties).  The vertical bars correspond to binomial
errors that include 68\% of the probability (1 $\sigma$ for Gaussian errors).
The data are consistent with no dependency with age with a rate of $\sim$ 20\%.
The data seems to suggest an overall decrease but so far is statistically not
significant, and if present may be due to an observational bias (because of the
deficiency of excesses around the K stars in the oldest age bin).  The number
of stars in the bins are (from young to old): 24, 57, 60, 52, 33, and 7 (the
high value of the oldest bin may be a small number statistical anomaly). For
comparison, A-type stars evolve on timescales of 400 Myr (Su et al. 2006).
Figure adapted 
from Trilling et al.  (2008) with data from Beichman et al.  (2006b).  }
\label{trilling13}       
\end{center}
\end{figure}

\begin{figure}[]
\begin{center}
\includegraphics[width=0.9\textwidth]{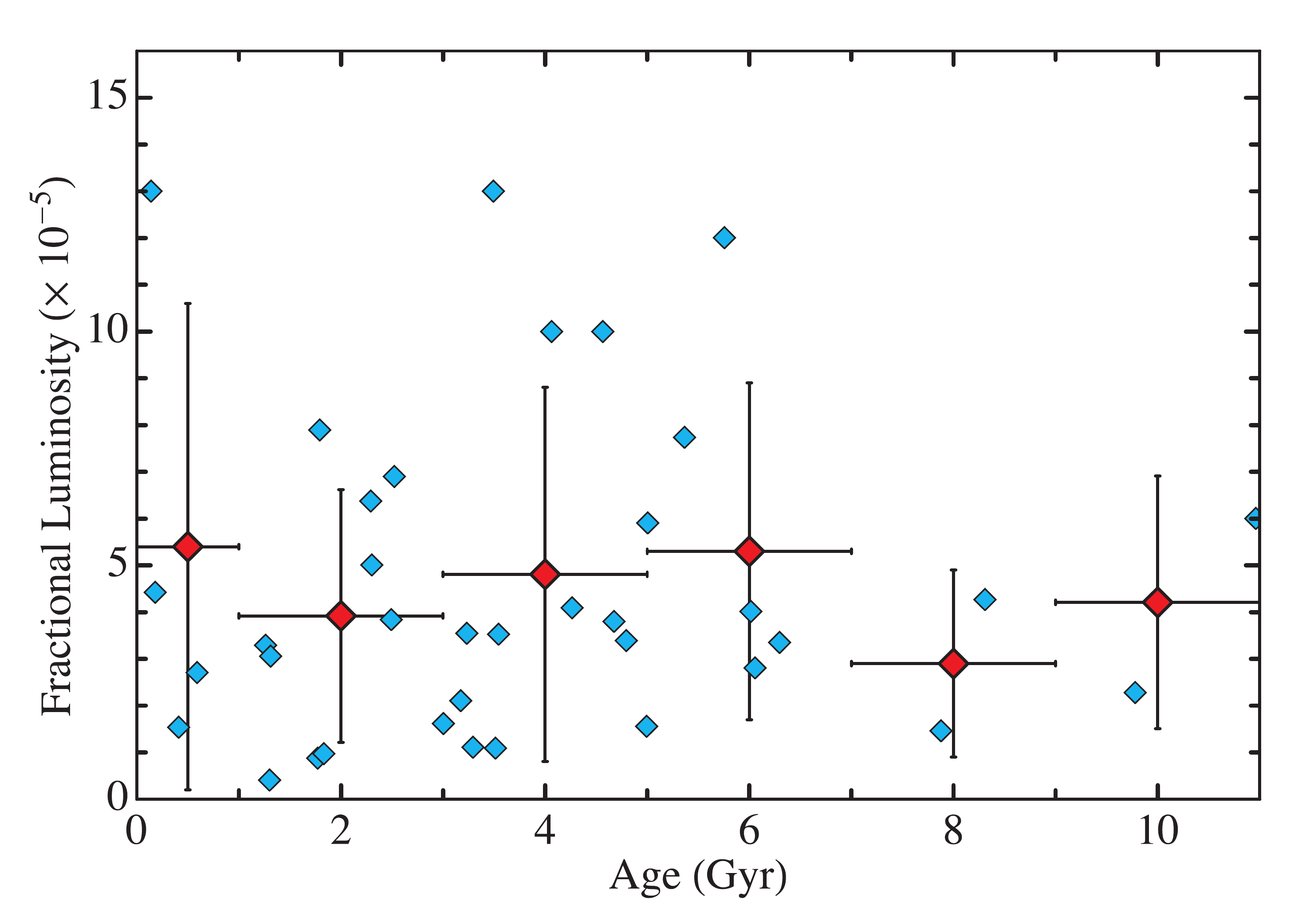}
\caption{The fractional luminosity of the debris disks,  L$_{dust}$/L$_{star}$,
as a function of age for FGK stars. The open symbols show the means within each
age bin; the horizontal and vertical error bars show the bin widths and 1
$\sigma$ errors, respectively. The data are consistent with no trend of
L$_{dust}$/L$_{star}$ with age, but there seems to be a deficiency of disks
with high L$_{dust}$/L$_{star}$ older than 6 Gyr. For comparison, the Solar
system is 4.5 Gyr old and is expected to have a dust disk with
L$_{dust}$/L$_{star}$ $\sim$ 10$^{-7}$--10$^{-6}$. Figure adapted 
from Trilling et al.  (2008) with data from Beichman et al. (2006b).  }
\label{trilling14}       
\end{center}
\end{figure}

\begin{figure}[]
\begin{center}
\includegraphics[width=0.8\textwidth]{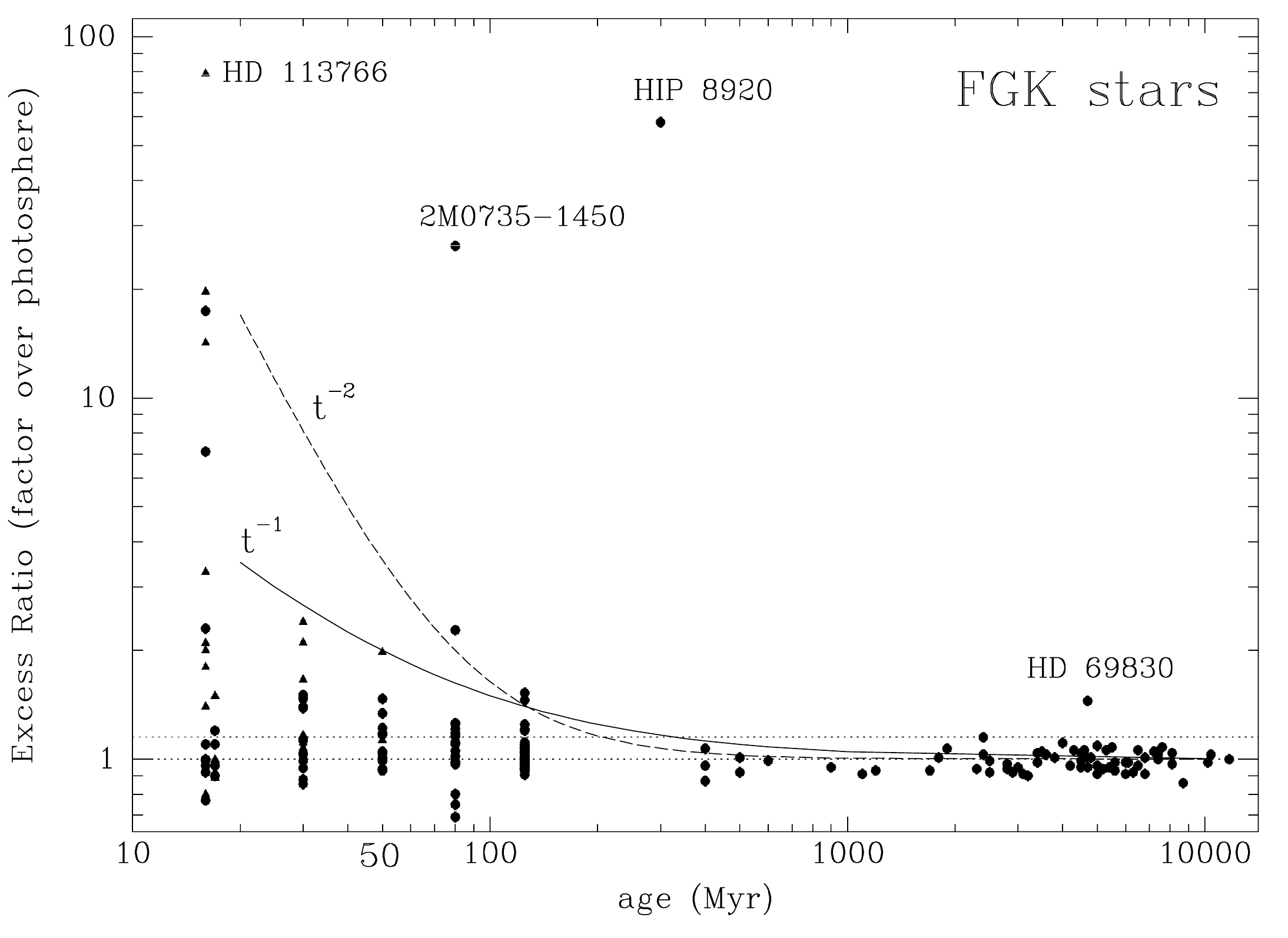}
\caption{Ratio of the 24 $\mu$m~Excess Emission over the expected stellar value
for FGK stars as a function of stellar age ({\it triangles} for F0--F4 stars
and {\it circles} for F5--K7 stars). The vertical alignments correspond to
stars in clusters or associations. The data agree broadly with collisional
cascade models of dust evolution (resulting in a 1/$t$ decay for the dust mass)
punctuated by peaks of dust production due to individual collisional events.
A-type stars show a similar behavior. Figure from Siegler et al. (2007) using
data from Gorlova et al. (2004), Hines et al.  (2006) and Song et al. (2005).
}
\label{siegler}       
\end{center}
\end{figure}

The {\it FEPS} survey of 328 FGK stars found that at  24 $\mu$m, the frequency of excess 
($>$ 10.2\% over the stellar photosphere) decreases from 14.7\% at ages $<$ 300 Myr 
 to 2\% for older stars; at 70 $\mu$m, there is no apparent dependency of the excess 
 frequency with stellar age, however, the amplitude of the 70 $\mu$m excess emission 
 seems to decline from stars 30--200 Myr in age to older stars (Hillenbrand et al. 2008 and 
 Carpenter et al. in preparation). 
 Figures \ref{trilling13} and \ref{trilling14} from Trilling et al. (2008) show
that for FGK type stars the debris disks incidence and fractional luminosity do
not have a strong dependency with stellar age in the 1--10 Gyr time frame, in
contrast with the 100--400 Myr evolution timescale of young (0.01--1 Gyr) stars
seen in Figure \ref{siegler}. Trilling et al. (2007) argues that this data suggests that the dominant physical
processes driving the evolution of the dust disks in young stars might be
different from those in more mature stars, and operate on different timescales:
while the former might be dominated by the production of dust during transient
events like the LHB in the Solar system or by individual collisions of large
planetesimals (like the one giving rise to the formation of the Moon), the
later might be the result of a more steady collisional evolution of a large
population of planetesimals. The debris disks evolution
observed by {\it Spitzer} for solar-type (Figure \ref{siegler} -- Siegler et
al. 2007) and A-type stars (Rieke et al.  2005 and Su et al. 2006)  indicate
that both transient and more steady state dust production processes play a
role; however, their relative importance and the question of how the dust
production could be maintained in the oldest disks for billions of years is
still under discussion. 

\begin{figure}[]
\begin{center}
\includegraphics[width=0.9\textwidth]{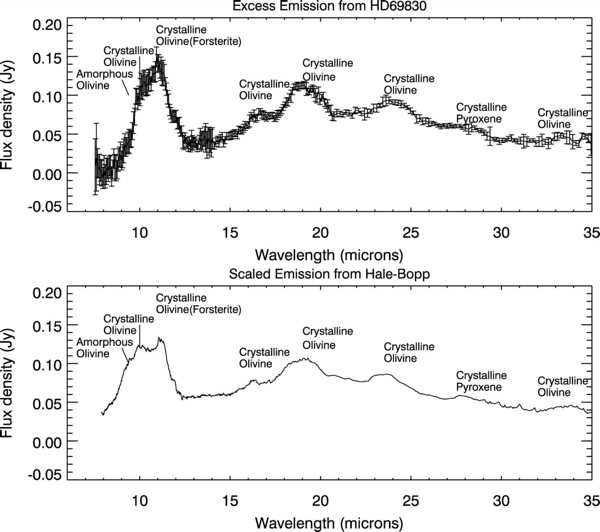}
\caption{Spectrum of the dust excess emission from HD 69830 (Beichman et al.
2005 -- {\it top}) compared to the spectrum of comet Hale-Bopp normalized to a
blackbody temperature of 400 K (Crovisier et al. 1996 -- {\it bottom}).  HD
69830 is one of the outliers in Figure \ref{siegler}.}
\label{hd69830}       
\end{center}
\end{figure}

An interesting example is HD 69830, one of the outliers in Figure
\ref{hd69830}, a KOV star (0.8 M$_{\odot}$, 0.45 L$_{\odot}$)  known to harbor
three Neptune-like planets inside 0.63 AU. It shows a strong excess at 24 $\mu$m
but no emission at 70 $\mu$m, indicating that the dust is warm and is located
close to the star. The spectrum of the dust excess (Figure \ref{hd69830}) shows
strong silicate emission lines thought to arise from small grains of 
highly processed material similar to that of a disrupted P- or D-type asteroid 
plus small icy grains,  likely located outside the outermost planet (Lisse et al. 2007). 
The observed levels of dust production are too high to be sustained for
the entire age of the star, indicating that the dust production processes are
transient (Wyatt et al. 2007).

\begin{figure}[]
\begin{center}
\includegraphics[width=0.9\textwidth]{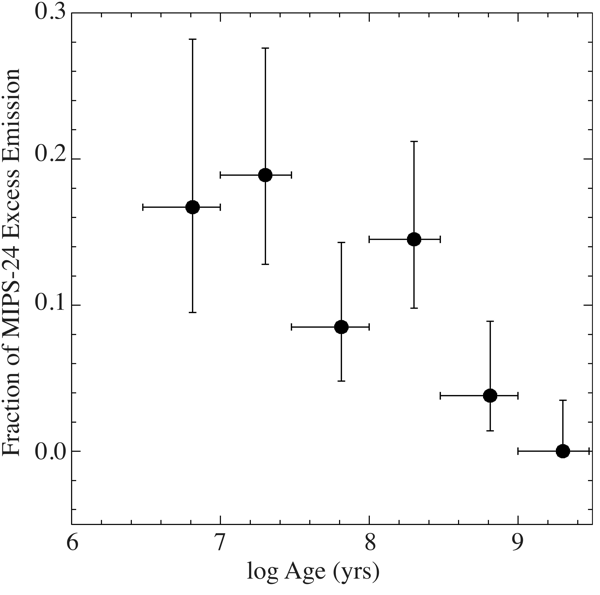}
\caption{Fraction of FGK type stars with 24 $\mu$m excess emission as a
function of stellar age from a sample of 328 stars. The data points correspond
to average values within a given age bin: (5/30) for stars 3--10 Myr, (9/48) for
10--30 Myr, (5/59) for 30--100Myr, (9/62) for 100--300Myr, (2/53) 300--1000 Myr.
The widths of the age bins are shown by the horizontal bars, while the
vertical bars show Poisson errors. Figure from Meyer et al. (2008).}
\label{meyer}       
\end{center}
\end{figure}

Whether the disks are transient or the result of the steady erosion of
planetesimals is of critical importance for the interpretation of the
statistics of the incidence rate of 24 $\mu$m excesses. For solar type stars,
the 24 $\mu$m emission traces the 4--6 AU region. Terrestrial planet formation
is expected to result in the production of large quantities of dust in this
region, due to gravitational perturbations produced by large 1000 km-sized
planetesimals that excite the orbits of a swarm of 1--10 km-size planetesimals,
increasing their rate of mutual collisions and producing dust (Kenyon \&
Bromley 2005). This warm dust can therefore serve as a proxy of terrestrial
planet formation. Figure \ref{meyer} show the frequency of 24 $\mu$m emission
for solar type (FGK) stars as a function of stellar age. This rate is $<$20\%
inside each age bin and decreases with age. If the dust-producing events are
very long-lived, the stars that show dust excesses in one age bin will also
show dust excesses at later times. In this case the frequency of warm dust
(which indirectly traces the frequency of terrestrial planet formation) is  $<$
20\%.  However, if the dust-producing events are short-lived, shorter than the
age bins, the stars showing excesses in one age bin are not the same as the
stars showing excesses at other age bins, i.e. they can produce dust at
different epochs, and in this case the overall frequency of warm dust is
obtained from adding all the frequencies in all age bins, which results in $>$
60\% (assuming that each star only has one epoch of high dust production). If
this is the case, the frequency of terrestrial planet formation would be high
(Meyer et al.  2008). However, the interpretation of the data would change if 
the  observed 24 $\mu$m excesses arise from the steady erosion of 
cold-KB-like disks (Carpenter et al., in preparation).  Spatially resolved observations 
able to directly locate the dust would help resolve this issue. \\
\\

\begin{svgraybox}\\
\textbf{Next generation} facilities will offer high-sensitivity,
high-resolution, multi-wavelength observations that should result in
major breakthroughs in the study of debris disks. Debris disks are proxies for the presence of planetesimals around mature stars
having a wide diversity of stellar types (A--K), suggesting that planetesimal
formation is a robust process. The study of the warm dust can tell us about the
frequency of terrestrial planet formation and the presence of asteroid-like
bodies, while the study of the cold dust sheds light on the population of
small bodies in KB-like regions. In addition, the study of debris disks around
stars having a wide diversity of ages can help us learn about the evolution of
planetary systems. However, the statistics so far are limited to dust disks
100--1000 $\times$ more luminous than that of our Solar system and the
observations are generally spatially unresolved. 
High-sensitivity observations with future telescopes like
{\it Herschel}, {\it JWST} and {\it ALMA} will be able to detect dust at the Solar system
level, will help us improve our understanding of the frequency of planetesimals
and, together with the result from planet searches, will show the diversity of
planetary systems. Multi-wavelength observations are critical to help
locate the dust in spatially unresolved disks and fundamental to interpret the
debris disks statistics. High-resolution imaging observations are very important
to directly locate the dust (circumventing the SED degeneracy), and to study
the structure of the debris disks, perhaps serving as a planet-detection method
sensitive to long-period Neptune-like planets that otherwise may be
undetectable in the foreseeable future. Multi-wavelength observations also play
a critical role in the interpretation of the structure because different
wavelengths trace different particle sizes which have distinct dynamical characters that
affect the disk morphology. 
\end{svgraybox}

\subsection{Photospheric Pollution}

Dust produced collisionally in a Kuiper belt may spiral to the central star
under the action of radiation and/or plasma drag, contaminating the photosphere
with metal-enriched material.  Separately, gravitational interactions and
dynamical instabilities in a Kuiper belt may eject large objects (comets),
causing some to impact the central star.  Both processes operate in our Solar
system but neither produces a spectrally distinctive signature, for example a
metal enrichment, on the Sun.  This is simply because the photosphere of the
Sun already contains a large mass of metals and the addition of dust or
macroscopic bodies makes only a tiny, fractional contribution.  

However, the atmospheres of some white dwarf stars offer much more favorable
opportunities for the detection of a photospheric pollution signal.  First,
many white dwarfs are naturally depleted in metals as a result of sedimentation
of heavy elements driven by their strong gravitational fields.  Since their
atmospheres should be very clean, quite modest masses of heavy-element
pollutants can be detected.  About one fifth of white dwarfs expected to have
pure hydrogen or pure helium atmospheres in fact show evidence for heavier
elements, most likely due to pollution from external sources.  Second, stellar
evolution leading to the white dwarf stage includes the loss of stellar mass
through an enhanced wind.  As the central star mass decreases, the semimajor
axes of orbiting bodies should increase, leading to dynamical instabilities
caused by resonance sweeping and other effects (Debes and Sigurdsson 2002).   
Separately, white dwarf stars with Oort clouds should experience a steady flux
of impacts from comets deflected inwards by stellar and galactic perturbations
(Alcock et al. 1986).

An observed depletion of carbon relative to iron may suggest that the infalling
material is cometary rather than of interstellar origin (Jura 2006).  

\subsection{Thermal Activation}

The blackbody temperature, in Kelvin, of an object located at distance,
$R_{AU}$, from a star of luminosity, $L_{\star}/L_{\odot}$, is $T_{BB}$ =  278
$R_{AU}^{-1/2} (L_{\star}/L_{\odot})^{1/4}$.  In the Kuiper belt today, at
$R_{AU}$ = 40, the blackbody temperature is $T_{BB}$ = 44 K.  Water sublimation
at this low temperature is negligible.   However, the sublimation rate is an
exponential function of temperature and, by $T_{BB}$ = 200 K, water sublimates
rapidly (with a mass flux $\sim$10$^{-4}$ kg m$^{-2}$ s$^{-1}$, corresponding
to ice recession at about 3 m yr$^{-1}$ for density 1000 kg m$^{-3}$).   A 1 km
scale body would sublimate away in just a few centuries.  By the above
relation, temperatures of 200 K are reached when $L_{\star}/L_{\odot}$ = 400,
for the same 40 AU distance.   

Stellar evolution into the red giant phase will drive the Sun's luminosity to
exceed this value after about 10 Gyr on the main-sequence, with an increase in
the luminosity (by up to a factor of $\sim$10$^4$), and in the loss of mass
through an enhanced stellar wind (from the current value, $\sim$10$^{-11}$
M$_{\odot}$ yr$^{-1}$, to $\sim$10$^{-7}$ M$_{\odot}$ yr$^{-1}$, or more).
When this happens, the entire Kuiper belt will light up as surface ices
sublimate and dust particles, previously embedded in the KBOs, are ejected into
space.  Not all KBOs will be destroyed by roasting in the heat of the giant
Sun: observations of comets near the Sun show that these bodies can insulate
themselves from the heat by the development of refractory mantles, consisting
of silicate and organic-rich debris particles that are too large to be ejected
by gas drag.  Still, the impact of the red giant phase should be dramatic and
suggests that the sublimated Kuiper belts of other stars might be detected
around red-giants.   The key observational signatures would be the thermal
excess itself, at temperatures appropriate to Kuiper belt-like distances, and
ring-like morphology.  Sensitivity at thermal wavelengths combined with high
angular resolution will lend \textit{JWST} to this type of observation, although the
overwhelming signal from the star itself will present a formidable
observational limitation to any imaging studies.

Water vapor has been reported around carbon stars that are not expected to show
water and interpreted as produced by sublimated comets (Ford and Neufeld 2001).
In IRC +10216, the mass of water is estimated as 3$\times$10$^{-5}$ M$_{\odot}$
(Melnick et al. 2001).  This is about 10 M$_{\oplus}$, or 100 times the mass of
the modern Kuiper belt but perhaps comparable to the mass of the Kuiper belt
when formed.   A possible explanation is that the water derives from sublimated
comets in an unseen Kuiper belt, but chemical explanations for this large water mass may also be possible
(Willacy 2004).  On the other hand, a search for (unresolved) thermal emission from dust around 66
first-ascent red giants proved negative, with limits on the Kuiper belt masses
of these stars near 0.1 M$_{\oplus}$, the current mass of the Kuiper belt (Jura 2004).

\begin{acknowledgement}
DJ was supported by a grant from NASA's Origins program, PL by an NSF Planetary Astronomy grant to DJ. A.M.M. is under contract with the Jet Propulsion Laboratory (JPL) funded by NASA through the Michelson Fellowship Program. A.M.M. is also supported by the Lyman Spitzer Fellowship at Princeton University. 
\end{acknowledgement}

\end{document}